\documentclass[aps,prc,twocolumn,showpacs,preprintnumbers,
               nofootinbib,float,superscriptaddress,longbibliography]{revtex4-1}
\usepackage{graphicx, fancybox}
\usepackage{amsmath,amssymb}
\usepackage[colorlinks=true, pdfstartview=FitV, linkcolor=red, citecolor=blue, urlcolor=blue]{hyperref}
\usepackage{color}
\usepackage{soul}
\usepackage{array}
\usepackage{url}
\usepackage[utf8]{inputenc}

\usepackage[normalem]{ulem}

\begin{document}

\title{A collision geometry-based 3D initial condition for relativistic heavy-ion collisions}

\author{Chun Shen}
\email{chunshen@wayne.edu}
\affiliation{Department of Physics and Astronomy, Wayne State University, Detroit, MI, 48201, USA}
\affiliation{RIKEN BNL Research Center, Brookhaven National Laboratory, Upton, NY 11973, USA}

\author{Sahr Alzhrani}
\email{sahr.alzhrani@wayne.edu}
\affiliation{Department of Physics and Astronomy, Wayne State University, Detroit, MI, 48201, USA}

\begin{abstract}
We present a simple way to construct 3D initial conditions for relativistic heavy-ion collisions based on the Glauber collision geometry. Local energy and momentum conservation conditions are imposed to set non-trivial constraints on our parameterizations of longitudinal profiles for the system's initial energy density and flow velocity. After calibrating parameters with charged hadron rapidity distributions in central Au+Au collisions, we test model predictions for particle rapidity distributions in d+Au and peripheral Au+Au collisions in the Beam Energy Scan (BES) program at Relativistic Heavy-Ion Collider (RHIC). Simulations and comparisons with measurements are also made for Pb+Pb collisions at Super Proton Synchrotron (SPS) energies.
We demonstrate that elliptic flow measurements in heavy-ion collisions at $\sqrt{s} \sim 10$ GeV can set strong constraints on the dependence of Quark-Gluon Plasma shear viscosity on temperature and net baryon chemical potential.
\end{abstract}

{\maketitle}

\section{Introduction}
\label{Intro}

Understanding the phase structure of QCD matter is one of the critical questions in relativistic heavy-ion physics.
First principles lattice QCD calculations have established that the transition from hadron resonance gas to the Quark-Gluon Plasma (QGP) phase at vanishing net-baryon density is a smooth cross over \cite{Aoki:2006we}. The presence of a first-order transition accompanied by a critical point at some finite net-baryon density has been conjectured based on many model calculations (see e.g., \cite{Stephanov:2004wx, Bzdak:2019pkr} for a review).
Current heavy-ion experiments in the RHIC BES program, the NA61/SHINE experiment at the Super Proton Synchrotron (SPS), as well as future experiments at the Facility for Antiproton and Ion Research (FAIR) and Nuclotron-based Ion Collider fAcility (NICA), produce hot and dense nuclear matter to probe an extensive temperature and baryon chemical potential region in the QCD phase diagram. Measurements from these collisions study the nature of the QCD phase transition from hadron gas to the QGP at different net baryon densities. Furthermore, the STAR Collaboration at the RHIC discovered non-zero global polarization of $\Lambda$ hyperons, which could indicate local fluid vorticity of $\omega \approx (9 \pm 1) \times 10^{21} s^{-1}$ \cite{STAR:2017ckg}. This result far surpasses the vorticity of all other known fluids in nature. 
Hence, it is a great interest to understand the origin of the RHIC $\Lambda$ polarization measurements.
Because the tiny fireballs created in heavy-ion collisions undergo ultra-fast yoctosecond dynamics, theoretical modeling of the system's multi-stage evolution is essential to elucidate physics from experimental measurements. In particular, phenomenological studies of precise flow measurements of the hadronic final states allow for the extraction of transport properties of the QGP in a baryon-rich environment \cite{Shen:2020gef}. 

Heavy-ion collisions at $\sqrt{s} \sim \mathcal{O}(10)$ GeV strongly violate longitudinal boost invariance and require full 3D modeling of their dynamics \cite{Shen:2017bsr}.
The geometry of incoming nuclei and the impact parameter between them determines the transverse shape of the produced fireball at a length scale of system size. Event-by-event fluctuations raise at multiple smaller length scales, such as nucleon-size fluctuations from random positions of nucleons inside the colliding nucleus \cite{Alver:2008aq} and sub-nucleon fluctuations from valence quarks and gluon fields \cite{Schenke:2012fw, Mantysaari:2016ykx, Mantysaari:2017cni}. Recently, there is a collective interest in understanding the longitudinal dynamics of heavy-ion collisions. 
The longitudinal profile of initial energy density is usually parameterized by a plateau-like function in symmetric collision systems \cite{Hirano:2005xf}. Variations of parametrizations were proposed to study the rapidity dependent directed flow as well as longitudinal flow fluctuations \cite{Bozek:2010vz, Bozek:2015bna, Broniowski:2015oif, Bozek:2018nne, Barej:2019xef}. 
Models that includes 3D dynamical evolution at pre-hydrodynamic phase are recently developed based on classical strings deceleration \cite{Denicol:2015nhu, Shen:2017bsr} and transport approaches \cite{Pang:2014pxa, Karpenko:2015xea, Du:2018mpf, Akamatsu:2018olk}. These models can provide a non-trivial correlation between the rapidity and space-time rapidity of energy-momentum and net baryon density currents \cite{Shen:2017fnn}. These models have predictive power for particle rapidity distributions at different collision energies.
In addition, there are new theoretical developments in exploring longitudinal flow observables with the 3+1D IP-Glasma model \cite{Schenke:2016ksl, McDonald:2020oyf}, understanding early-stage baryon stopping from Color Glass Condensate (CGC) effective theory based approaches in the fragmentation region \cite{Li:2018ini, McLerran:2018avb}, and from a holographic approach at intermediate couplings \cite{Attems:2018gou}.

In this work, we assume that all the energy and momentum from the two colliding nuclei are deposited into fluid dynamic fields in 3D hydrodynamic simulations. Rare processes, such as QCD jets and heavy-quark productions, carry negligible energy and momentum compared to those of the entire collision system. Using these constraints from total energy and net longitudinal momentum, we can reduce the number of model parameters in longitudinal profiles for the initial energy density. Specifically, we do not require an overall normalization factor for system's initial energy density profile anymore. By adjusting the width of the energy density plateau to match the measured charged hadron multiplicity at mid-rapidity, this model can make predictions for the centrality dependence of particle production as well as particle rapidity distributions. In this work, we find that imposing local energy-momentum conservation in 3D initial conditions can lead to a non-trivial transverse energy density profile. 
Our proposed scheme can straightforwardly extend to event-by-event simulations in the future. Because this approach ignored pre-equilibrium dynamics during the two colliding nuclei pass through each other, our results serve as a seline for future comparisons with more realistic dynamical initialization frameworks. 

In the next section, we will describe the procedure of imposing local energy-momentum conservation and our parameterization of longitudinal profiles. Using event-averaged density profiles for nuclear thickness functions, we will analyze the longitudinal distribution and collision centrality dependence of the fireball's initial eccentricity coefficients. Sec.~\ref{sec:results} will present our theoretical results from evolving 3D initial conditions with hydrodynamics + hadronic transport hybrid framework. We will present a phenomenological study compared with experimental measurements at RHIC BES and SPS programs, focusing on charged hadron and identified particle rapidity distributions. The particle's directed and elliptic flow coefficients will be studied as well. Sec.~\ref{sec:conclusion} is devoted to some concluding remarks.

\section{The Model Framework} \label{sec:model}

In this section, we discuss an extension of geometry-based 2D initial conditions to 3D with parametric longitudinal profiles for energy and net baryon densities. Conservation laws of local energy and momentum are imposed to constrain our parametrization.

\subsection{Local energy-momentum conservation}

The nucleus thickness function determines collision geometry in the transverse plane,
\begin{eqnarray}
    T_{A(B)}(x, y) &=& \sum_{i \in \mathrm{participants}} \frac{1}{2 \pi \sigma_\perp^2} \nonumber \\
    && \quad \times \exp\left[ -\frac{(x - x_i)^2}{2\sigma_\perp^2} - \frac{(y - y_i)^2}{2\sigma_\perp^2}\right].
\end{eqnarray}
We denote the colliding nucleus $A$ as the projectile, which has a positive velocity along the longitudinal direction, and the nucleus $B$ to fly to negative a direction as the target. The impact parameter vector $\vec{b}$ is defined to start from the target to the projectile. Individual nucleons are assumed to have a Gaussian profile in the transverse plane with a parametric width of $\sigma_\perp$. Before the collision, all the nucleons fly with the beam rapidity determined by the center-of-mass collision energy $\sqrt{s_\mathrm{NN}}$,
\begin{equation}
    y_\mathrm{beam} = \mathrm{arccosh} \left(\frac{\sqrt{s_\mathrm{NN}}}{2 m_N} \right).
\end{equation}
Here the nucleon mass is $m_N$. With nucleus thickness functions, we can determine the local collision energy and net longitudinal momentum at any point in the transverse plane $(x, y)$,
\begin{eqnarray}
    E(x, y) &=& [T_A(x, y) + T_B(x, y)] m_N \cosh(y_\mathrm{beam}) \nonumber \\
    &\equiv& M(x, y) \cosh(y_\mathrm{CM}(x, y)) \\
    P_z(x, y) &=& [T_A(x, y) - T_B(x, y)] m_N \sinh(y_\mathrm{beam}) \nonumber \\
    &\equiv& M(x, y) \sinh(y_\mathrm{CM}(x, y)).
\end{eqnarray}
We can define a local invariant mass $M(x, y)$ and a center-of-mass rapidity $y_\mathrm{CM}(x, y)$,
\begin{eqnarray}
    y_\mathrm{CM}(x, y) &=& \mathrm{arctanh}\left[\frac{T_A - T_B}{T_A + T_B} \mathrm{tanh}(y_\mathrm{beam}) \right] \label{eq:ycm} \\
    M(x, y) &=& m_N \sqrt{T_A^2 + T_B^2 + 2 T_A T_B \cosh(2y_\mathrm{beam})}. \label{eq:Minv}
\end{eqnarray}
These conditions will be used to constrain the longitudinal profiles at matching between initial conditions and hydrodynamic fields. It is worth noting that the factor $\cosh(2y_\mathrm{beam}) \gg 1$ for $\sqrt{s_\mathrm{NN}} > 5$\,GeV. At these collision energy, the local invariant mass at a transverse point $(x, y)$ is proportional to $\sqrt{T_A(x, y) T_B(x, y)}$. 

At the beginning of hydrodynamic simulations, we consider hydrodynamic initial conditions on a constant proper time surface $\tau \equiv \sqrt{t^2 - z^2} = \tau_0$,
\begin{eqnarray}
     && M(x, y) \cosh[y_\mathrm{CM}(x, y)] = \int d^3 \Sigma_\mu T^{\mu t}(x, y, \eta_s) \nonumber \\
    && \qquad  = \int \tau_0 d \eta_s (T^{\tau \tau}(x, y, \eta_s) \cosh(\eta_s) \nonumber \\
    && \qquad \qquad \qquad \qquad + \tau_0 T^{\tau \eta}(x, y, \eta_s) \sinh(\eta_s)) \label{eq:rule1} \\
    && M(x, y) \sinh[y_\mathrm{CM}(x, y)] = \int d^3 \Sigma_\mu T^{\mu z}(x, y, \eta_s) \nonumber \\
    && \qquad = \int \tau_0 d \eta_s (T^{\tau \tau}(x, y, \eta_s) \sinh(\eta_s) \nonumber \\
    && \qquad \qquad \qquad \qquad + \tau_0 T^{\tau \eta}(x, y, \eta_s) \cosh(\eta_s)) \label{eq:rule2}.
\end{eqnarray}
Here $T^{\tau \tau}(x, y, \eta_s)$ and $T^{\tau \eta}(x, y, \eta_s)$ are components of the system's energy-momentum tensor at $\tau = \tau_0$. We can rewrite Eqs.~(\ref{eq:rule1}) and (\ref{eq:rule2}) as follows,
\begin{eqnarray}
    M(x, y) &=& \int \tau_0 d \eta_s (T^{\tau \tau} \cosh(\eta_s - y_\mathrm{CM}) \nonumber \\
    && \qquad \qquad + \tau_0 T^{\tau \eta} \sinh(\eta_s - y_\mathrm{CM})) \\
    0 &=& \int \tau_0 d \eta_s (T^{\tau \tau} \sinh(\eta_s - y_\mathrm{CM}) \nonumber \\
    && \qquad \qquad + \tau_0 T^{\tau \eta} \cosh(\eta_s - y_\mathrm{CM})).
\end{eqnarray}
These two equations suggest that $T^{\tau\tau}(\eta_s)$ should be an even function to $y_\mathrm{CM}$ and $\tau_0 T^{\tau \eta}(\eta_s)$ is odd  w.r.t $y_\mathrm{CM}$.
In general, when we choose longitudinal profiles for local energy density and flow velocity profiles, we need to ensure conservation of energy and momentum in Eqs.~(\ref{eq:rule1}) and (\ref{eq:rule2}). These two conditions can eliminate two degrees of freedom in the model parameter space.

\subsection{A simple parametrization of longitudinal profiles for local energy density and flow velocity}

To keep our model simple, we assume the Bjorken flow for system's flow velocity profile at $\tau = \tau_0$,
\begin{equation}
    u^\mu (x, y, \eta_s) = (\cosh(\eta_s), 0, 0, \sinh(\eta_s)).
\end{equation}
This assumption significantly reduces the complexity of system's initial energy-momentum tensor, making it only a function of the local energy density, $T^{\tau\tau} (x, y, \eta_s) = e(x, y, \eta_s)$ and $T^{\tau \eta} = 0$. Energy-momentum conservation leads to
\begin{eqnarray}
    M(x, y) &=& \int \tau_0 d \eta_s e(x, y, \eta_s) \cosh(\eta_s - y_\mathrm{CM}) \\
    0 &=& \int \tau_0 d \eta_s e(x, y, \eta_s) \sinh(\eta_s - y_\mathrm{CM}).
\end{eqnarray}
We choose a symmetric rapidity profile parameterization w.r.t $y_\mathrm{CM}$ for the local energy density \cite{Hirano:2005xf},
\begin{eqnarray}
    && e (x, y, \eta_s; y_\mathrm{CM})  = \mathcal{N}_e(x, y) \nonumber \\
    && \!\!\!\!\!\quad \times \exp\left[- \frac{(\vert \eta_s - y_\mathrm{CM}\vert  - \eta_0)^2}{2\sigma_\eta^2} \theta(\vert \eta_s - y_\mathrm{CM} \vert - \eta_0)\right].
    \label{eq:eprof}
\end{eqnarray}
Here the parameter $\eta_0$ determines the width of the plateau and the $\sigma_\eta$ controls how fast the energy density falls off at the edge of the plateau. In a highly asymmetric situation $T_A(x, y) \gg T_B(x, y)$, the center-of-mass rapidity $y_\mathrm{CM}(x, y) \rightarrow y_\mathrm{beam}$. To make sure there is not too much energy density deposited beyond the beam rapidity, we set $\eta_0(x, y) = \mathrm{min}(\eta_0, y_\mathrm{beam} - y_\mathrm{CM}(x, y))$.
The normalization factor $\mathcal{N}_e (x, y)$ is not a free parameter in our model. It is determined by the local invariant mass $M(x, y)$,
\begin{eqnarray}
   &&\mathcal{N}_e (x, y) = \frac{M(x, y)}{2 \sinh(\eta_0) + \sqrt{\frac{\pi}{2}} \sigma_\eta e^{\sigma_\eta^2/2} C_\eta}  \\
   &&C_\eta = e^{\eta_0}\mathrm{erfc}\left(-\sqrt{\frac{1}{2}} \sigma_\eta\right)  + e^{-\eta_0}\mathrm{erfc}\left(\sqrt{\frac{1}{2}} \sigma_\eta\right).
\end{eqnarray}
Here $\mathrm{erfc}(x)$ is the complementary error function. Because the local invariant mass $M(x, y) \propto \sqrt{T_A(x, y) T_B(x, y)}$, our choice of the longitudinal profile leads to local energy density $e(x, y) \propto \sqrt{T_A(x, y) T_B(x, y)}$ inside the plateau region. At the top RHIC and LHC energy, the plateau width $\eta_0$ is large. Our derivation suggests that the transverse energy density in the mid-rapidity region should scales with $\sqrt{T_A(x, y) T_B(x, y)}$ instead of participant nucleon profile or binary collision profile. The scaling with $\sqrt{T_A(x, y) T_B(x, y)}$ was preferred from the current state-of-the-art Bayesian extraction \cite{Bernhard:2016tnd, Bernhard:2019bmu}. Our model provides a physics explanation that this scaling is rooted from conservation of energy and longitudinal momentum.  

We prepare event-averaged nuclear thickness functions based on the Monte-Carlo Glauber model using an open source code package \texttt{superMC} within the \texttt{iEBE-VISHNU} framework \cite{Shen:2014vra, superMC}. Centrality classes are determined with the number of participating nucleons in each collision.
Over 10,000 fluctuating initial conditions in every centrality bin are averaged after rotating each event by its second order participant plane angle to align the respective eccentricities \cite{Hirano:2009ah, Qiu:2011iv}. 

We use the same net baryon density profile as was in Ref.~\cite{Denicol:2018wdp}, 
\begin{eqnarray}
    && f^{A}_{n_B} (\eta_s) = \mathcal{N}_{n_B} \left\{\theta(\eta_s - \eta_{B,0}) \exp\left[ - \frac{(\eta_s - \eta_{B,0})^2}{2\sigma_{B, \mathrm{out}}^2} \right] \right. \nonumber \\
    && \qquad \qquad + \theta(\eta_{B,0} - \eta_s) \exp\left[ - \frac{(\eta_s - \eta_{B,0})^2}{2\sigma_{B, \mathrm{in}}^2} \right] \bigg\}
    \label{eq:nBprofr}
\end{eqnarray}
and
\begin{eqnarray}
    && f^{B}_{n_B} (\eta_s) = \mathcal{N}_{n_B} \bigg\{\theta(\eta_s + \eta_{B,0}) \exp\left[ - \frac{(\eta_s + \eta_{B,0})^2}{2\sigma_{B, \mathrm{in}}^2} \right] \nonumber \\
    && \qquad \qquad + \theta(- \eta_{B,0} - \eta_s) \exp\left[ - \frac{(\eta_s + \eta_{B,0})^2}{2\sigma_{B, \mathrm{out}}^2} \right] \bigg\}.
    \label{eq:nBprofl}
\end{eqnarray}

\begin{table}[ht!]
    \centering
    \begin{tabular}{|c|c|c|c|c|c|c|}
    \hline
        $\sqrt{s_\mathrm{NN}}$ (GeV) & $\tau_0$ (fm/$c$) & $\eta_0$ & $\sigma_\eta$ & $\eta_{B,0}$ & $\sigma_{B, \mathrm{in}}$ & $\sigma_{B, \mathrm{out}}$ \\ \hline
        AuAu \& dAu @ 200     &   1.0     &   2.5     &   0.6 &   3.5     &   2.0     &   0.1 \\ \hline
        AuAu \& dAu @ 62.4    &   1.0     &   2.25    &   0.3 &   2.7     &   1.9     &   0.2 \\ \hline
        AuAu \& dAu @ 39      &   1.3     &   1.9     &   0.3 &   2.2     &   1.6     &   0.2 \\ \hline
        AuAu@27      &   1.4     &   1.6    &   0.3 &   1.8     &   1.5     &   0.2 \\ \hline
        AuAu \& dAu @ 19.6    &   1.8     &   1.3     &   0.3 &   1.5     &   1.2     &   0.2 \\ \hline
        AuAu@14.5    &   2.2    &   1.15    &   0.3 &   1.4     &   1.15    &   0.2 \\ \hline
        AuAu@7.7     &   3.6     &   0.9     &   0.2 &   1.05    &   1.0     &   0.1 \\ \hline \hline
        PbPb@17.3    &   1.8     &   1.25    &   0.3 &   1.6 &   1.2    &   0.2 \\ \hline
        PbPb@8.77    &   3.5     &   0.95    &   0.2 &   1.2    &   1.0     &   0.1 \\ \hline
    \end{tabular}
    \caption{The model parameters for longitudinal envelope profiles for system's local energy density and net baryon density.}
    \label{table1}
\end{table}

\begin{figure}[t!]
    \centering
    \includegraphics[width=1.0\linewidth]{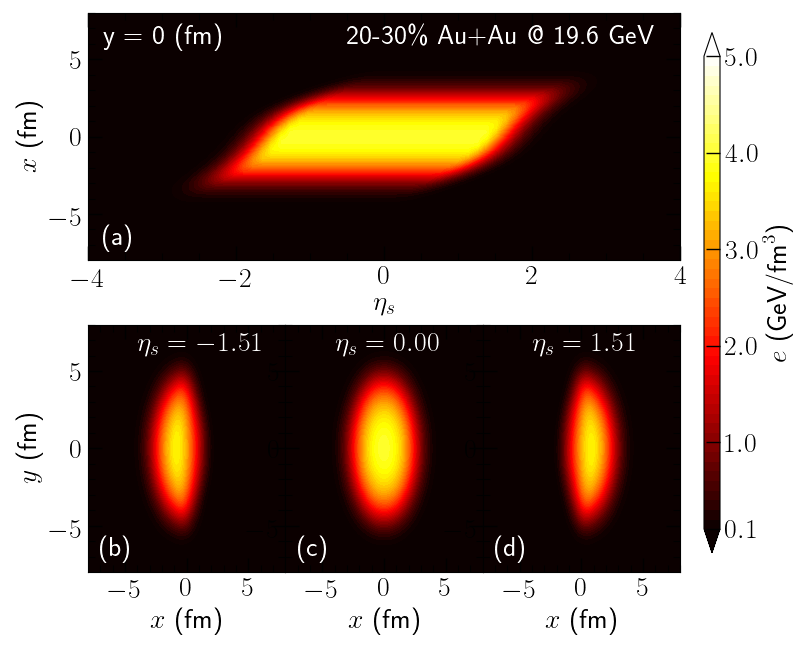}
    \caption{Color contour plot for the local energy density distribution at $\tau = 1.8$ fm/$c$ in 20-30\% Au+Au collisions at 19.6 GeV. }
    \label{fig:eprof}
\end{figure}

Table~\ref{table1} summaries the values of model parameters for the longitudinal profiles in Eqs.~(\ref{eq:eprof}), (\ref{eq:nBprofr}), and (\ref{eq:nBprofl}). They are calibrated with charged hadron and net proton rapidity distributions in central Au+Au collisions at different collision energies. With these parameters, we illustrate a 3D local energy density distribution for 20-30\% Au+Au collisions at $\sqrt{s_\mathrm{NN}} = 19.6$ GeV in Fig.~\ref{fig:eprof}.
Because the two colliding nuclei are offset by a finite impact parameter, their nuclear thickness functions at the edge of the overlapping region are not symmetric. This asymmetry shifts the local energy density plateau by a space-time rapidity $\Delta \eta_s = y_\mathrm{CM}(x, y)$ along the longitudinal direction. This effect makes a wider longitudinal distribution for energy density when collisions are asymmetric in the transverse plane. Figs.~\ref{fig:eprof}b-d further show that the transverse shape of energy density becomes more eccentric at forward and backward rapidity regions. 

\subsection{Boost-invariance breaking effects on fireballs' transverse geometry}

Before we perform hydrodynamic simulations, it is instructive to study the dependence of collision systems' initial eccentricities on centrality and space-time rapidity. The initial eccentricity coefficients can be defined as,
\begin{eqnarray}
    \vec{\mathcal{E}}_1 &\equiv& \varepsilon_1(\eta_s) e^{i \Psi_1(\eta_s)} = -\frac{\int d^2 r \tilde{r}^3 e^{i \tilde{\phi}}e(r, \phi, \eta_s)}{\int d^2 r \tilde{r}^3 e(r, \phi, \eta_s)} \\
    \vec{\mathcal{E}}_n &\equiv& \varepsilon_n(\eta_s) e^{i n \Psi_n(\eta_s)} \nonumber \\
    &=& -\frac{\int d^2 r \tilde{r}^n  e^{i n \tilde{\phi}}e(r, \phi, \eta_s) }{\int d^2 r \tilde{r}^n e(r, \phi, \eta_s)} \quad \mbox{for } (n \ge 2).
\end{eqnarray}
Here we compute those eccentricity coefficients with respect to the energy density weighted center-of-mass point $(x_0(\eta_s), y_0(\eta_s))$ in every space-time rapidity slice. The transverse radius and azimuthal angle are defined as $\tilde{r}(x, y, \eta_s) = \sqrt{(x - x_0(\eta_s))^2 + (y - y_0(\eta_s))^2}$ and $\tilde{\phi}(x, y, \eta_s) = \arctan[(y - y_0(\eta_s))/(x - x_0(\eta_s))]$. The center-of-mass point is computed using local energy density as a weight,
\begin{eqnarray}
    x_0(\eta_s) &=& \frac{\int d^2 r x e(r, \phi, \eta_s)}{\int d^2 r e(r, \phi, \eta_s)} \\
    y_0(\eta_s) &=& \frac{\int d^2 r y e(r, \phi, \eta_s)}{\int d^2 r e(r, \phi, \eta_s)}.
\end{eqnarray}
\begin{figure}[ht!]
    \centering
    \includegraphics[width=0.95\linewidth]{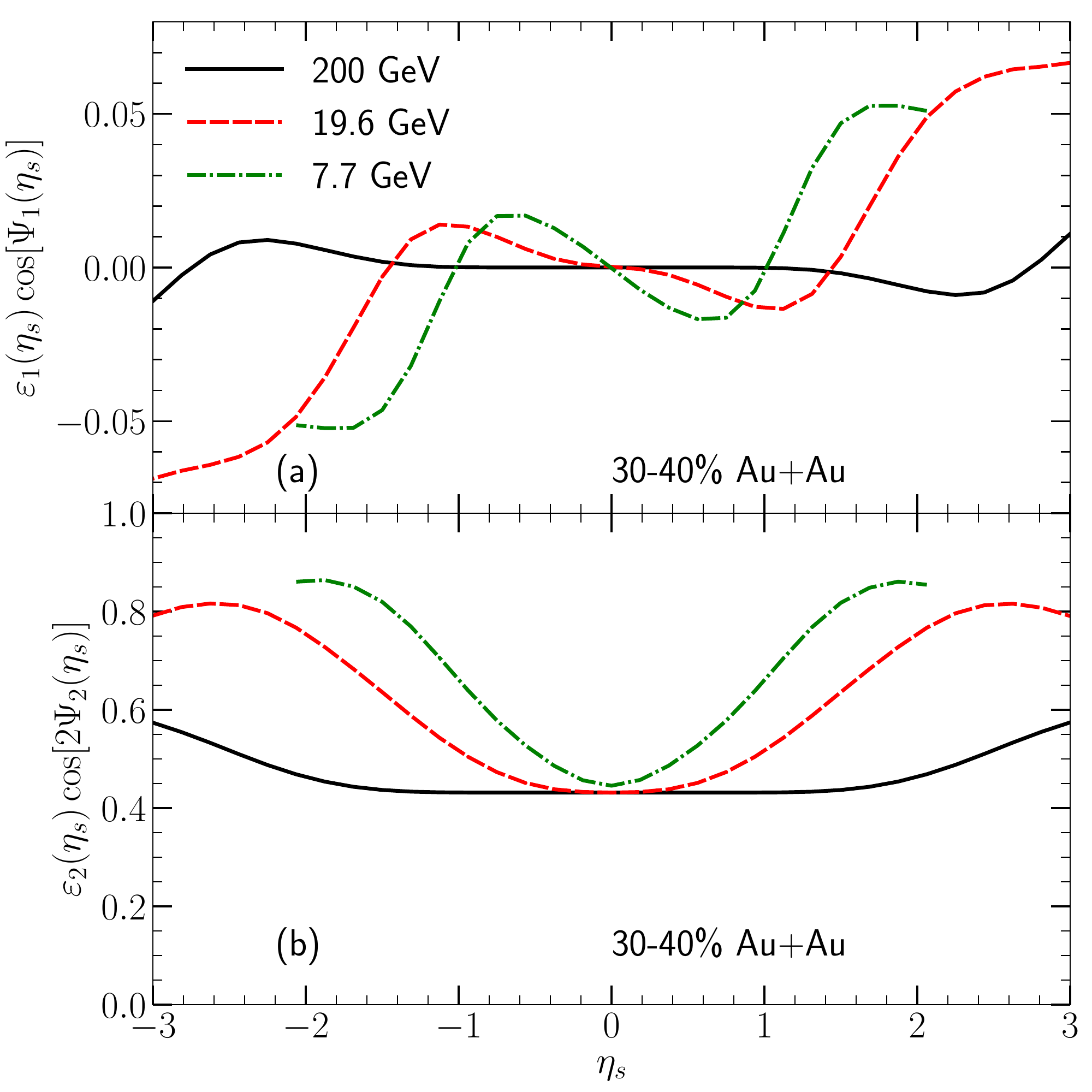}
    \caption{The space-time rapidity distribution of initial eccentricity coefficients $\varepsilon_1(\eta_s)$ and $\varepsilon_2(\eta_s)$ for 30-40\% Au+Au collisions at 200, 19.6, and 7.7 GeV.}
    \label{fig:eccn_rap}
\end{figure}

The space-time rapidity distribution of initial eccentricity vectors are demonstrated in Fig.~\ref{fig:eccn_rap} for 30-40\% Au+Au collisions at 7.7, 19.6, and 200 GeV. We set the impact parameter to point along the $+x$ direction so that the imaginary part of $\vec{\mathcal{E}}_n$ is zero. The dipole deformation vector $\vec{\mathcal{E}}_1(\eta_s)$ is an odd function in space-time rapidity, which reflects shape variations of transverse energy density profiles along the $\eta_s$ direction shown in Fig.~\ref{fig:eprof}. Collisions at lower energy show stronger dipole deformations. At 19.6 GeV, the dipole vector $\vec{\mathcal{E}}_1(\eta_s)$ points to the direction where the colliding nucleus sits in the transverse plane in a forward region with $\eta_s > 1.4$. In this region, the shape of local energy density is dominated by the projectile nuclear thickness function. In the meantime, the direction of $\vec{\mathcal{E}}_1(\eta_s)$ is opposite near the mid-rapidity region, $\vert \eta_s \vert < 1.4$. Figure~\ref{fig:eprof}d shows that the shape of the target nuclear thickness function generates the steepest gradient in the local energy density profile at $\eta_s \sim 1.5$. It flips the dipole vector's direction.

The initial $\vec{\mathcal{E}}_2(\eta_s)$ vector is an even function in space-time rapidity. At 200 GeV, the magnitude of elliptic deformation $\varepsilon_2(\eta_s)$ remains constant for a space-time rapidity window of $\eta_s \sim [-1.5, 1.5]$, showing approximated boost-invariance in this region. Such a plateau region shrinks quickly as collision energy decreases. The spatial ellipticity increases in forward and backward rapidity regions because of the shifts induced by the local net longitudinal momentum. This is consistent with those energy density profiles illustrated in Figs.~\ref{fig:eprof}b-d. 

\begin{figure}[ht!]
    \centering
    \includegraphics[width=1.0\linewidth]{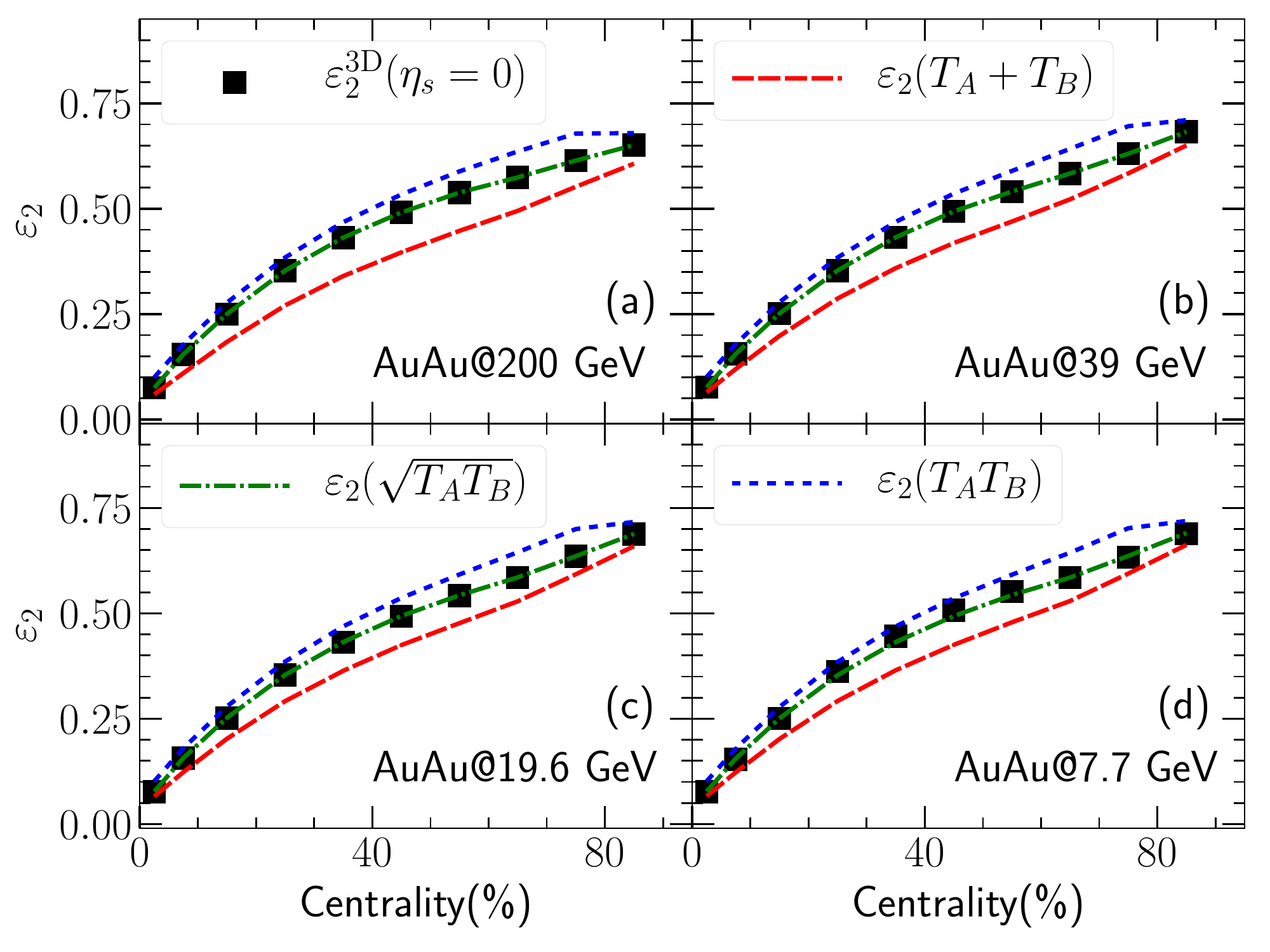}
    \caption{Centrality dependence of the initial eccentricity $\varepsilon_2$ for Au+Au collisions at 200, 39, 19.6, and 7.7 GeV. The mid-rapidity eccentricities from 3D energy density profiles are compared with those assuming longitudinal boost-invariant and local energy density being proportional to $T_A + T_B$, $\sqrt{T_A T_B}$, and $T_A T_B$. Here $T_A$ and $T_B$ are the nuclear thickness functions.}
    \label{fig:ecc2scaling}
\end{figure}
In Figure~\ref{fig:ecc2scaling}, we further study the centrality dependence of $\varepsilon_2$ at mid-rapidity in Au+Au collisions at four collision energies. They are compared with those eccentricities estimated with different functions of the two nuclear thickness function. The case $\varepsilon_2(T_A + T_B)$ corresponds to the ellipticity of the participant nucleon profile, while the results from $\varepsilon_2(T_AT_B)$ represent the quadruple deformation of the system's binary collision profile. Based on our derivation in the previous section, our eccentricity $\epsilon^\mathrm{3D}_2$ at $\eta_s = 0$ should approximately scale with the $\sqrt{T_A T_B}$ profile. With the model parameters listed in Table~\ref{table1}, we find the $\sqrt{T_A T_B}$ eccentricity scaling works well even at $\sqrt{s} = 7.7$\,GeV.

\subsection{Temperature and net baryon chemical potential dependent QGP specific shear viscosity}

The 3D energy density and net baryon density profiles serve as initial conditions for hydrodynamics and hadronic transport simulations. In this work, we use the open-source 3D viscous hydrodynamic code package \texttt{MUSIC} \cite{Schenke:2010nt, Schenke:2011bn, Paquet:2015lta, Denicol:2018wdp, MUSIC} to simulate fluid dynamical evolution of the collision system's energy, momentum, and net baryon density,
\begin{eqnarray}
\partial_\mu T^{\mu\nu} &=& 0 \\
\partial_\mu J^{\mu}_B &=& 0.
\end{eqnarray}
Because the overlapping time for the two colliding nuclei to pass through each other is not negligible at $\sqrt{s} \sim \mathcal{O}(10)$\,GeV \cite{Karpenko:2015xea, Shen:2017bsr}, we will start our hydrodynamic simulations at a $\tau = \tau_0 > \tau_\mathrm{overlap}$. The values of $\tau_0$ used at different collision energies are listed in Table~\ref{table1}. They are calibrated to reproduce the measured identified particle mean $p_T$ in Au+Au collisions. Hydrodynamic equations of motion are solved with a lattice QCD based equation of state (EoS) at finite net baryon density, NEoS-BQS \cite{Monnai:2019hkn, Monnai:2020pcw}. This EoS imposes strangeness neutrality and constrains the local net electric charge density to be 0.4 times of the local net baryon density. We include shear viscous effect during hydrodynamic simulations.
We explore a temperature and net baryon chemical potential dependent specific shear viscosity $(\eta/s)(T, \mu_B)$, which is parametrized as
\begin{equation}
    \frac{\eta}{s} (T, \mu_B) = \left(\frac{\eta}{s} \right)_0 f_T(T) \left(\frac{e+P}{T s}\right) f_{\mu_B} (\mu_B).
    \label{eq:eta_over_s}
\end{equation}
Here the temperature and net baryon chemical potential dependent $(\eta/s)(T, \mu_B)$ is defined via the following functions,
\begin{equation}
    f_T(T) = 1 + T_\mathrm{slope} \frac{T - T_c}{T_c - T_\mathrm{low}} \theta(T_c - T)
\end{equation}
and
\begin{equation}
    f_{\mu_B}(\mu_B) = 1 + \mu_{B,\mathrm{slope}} \left( \frac{\mu_B}{\mu_{B, \mathrm{scale}}} \right)^\alpha.
\end{equation}
In addition, the factor $(e+P)/(Ts)$ introduces some extra $T$ and $\mu$ dependence through the equation of state NEoS-BQS $(n_S = 0, n_Q = 0.4n_B)$ at finite densities,
\begin{eqnarray}
    \frac{e+P}{T s} &=& 1 + \sum_{i=B,Q,S} \frac{n_i}{s} \frac{\mu_i}{T} \nonumber \\
    &=& 1 + \frac{n_B}{s} \frac{\mu_B + 0.4\mu_Q}{T}.
\end{eqnarray}
\begin{figure}[ht!]
  \centering
  \includegraphics[width=1.0\linewidth]{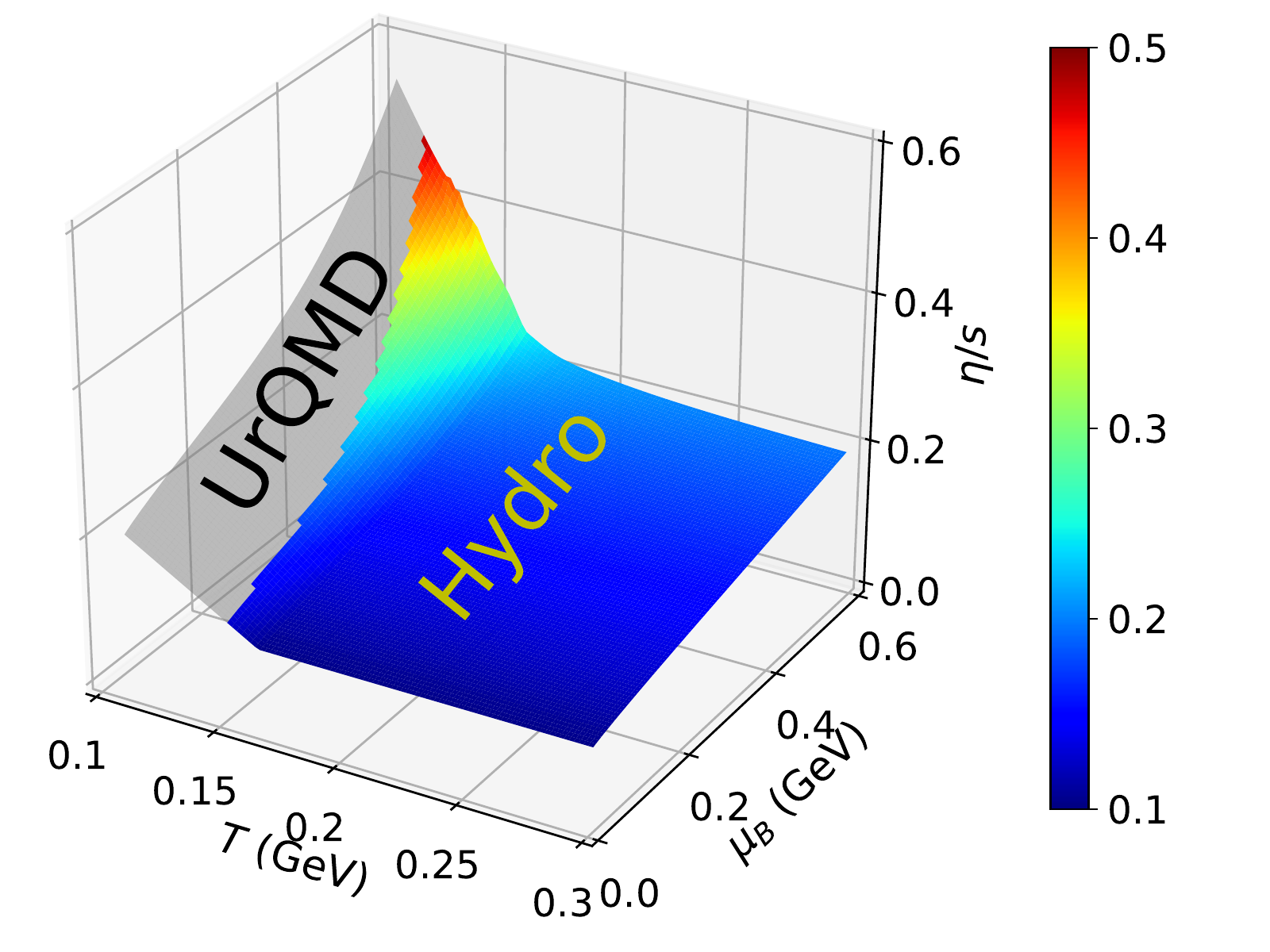}
  \caption{The specific shear viscosity $(\eta/s)(T, \mu_B)$ used in our hydrodynamic simulations. The grey area indicates the hadronic transport region with $e_\mathrm{sw} < 0.26$ GeV/fm$^3$.}
  \label{fig:vis}
\end{figure}

Calibrating our full hybrid simulations with charged hadron elliptic flow measurements at the RHIC BES program (see results in Sec.~\ref{sec:results:transflow} below), we find the following set of parameters, $(\eta/s)_0 = 0.1$, $T_c = 0.165$ GeV, $T_\mathrm{low} = 0.1$ GeV, $T_\mathrm{slope} = 1.2$, $\mu_{B,\mathrm{slope}} = 0.9$, $\mu_{B, \mathrm{scale}} = 0.6$ GeV, and $\alpha = 0.8$.

Figure~\ref{fig:vis} shows the temperature and net baryon chemical potential dependent specific shear viscosity $(\eta/s)(T, \mu_B)$ used in our hydrodynamic simulations. It increases rapidly at low temperature and large $\mu_B$ regions.
Our hydrodynamic simulations transition to a microscopic transport description, \texttt{UrQMD} \cite{Bass:1998ca,Bleicher:1999xi}, on a constant energy density hyper-surface with $e_\mathrm{sw} = 0.26$\,GeV/fm$^3$. The hadronic transport phase is indicated as the grey area in Fig.~\ref{fig:vis}.

\section{Results and Discussions}\label{sec:results}

In this section, we will present our model calibrations and test its predictions with measurements in the RHIC BES and the CERN SPS programs. Our results will serve as a baseline for future comparisons with more realistic event-by-event simulations with a dynamical initialization scheme \cite{Shen:2017bsr}.

\subsection{Particle productions and their rapidity distributions}

While the overall normalization of our longitudinal profile for energy density is constrained by system's collision energy in Eq.~(\ref{eq:eprof}), the widths of energy density plateaus at different $\sqrt{s}$ need to be calibrated with measured charged hadron pseudo-rapidity distributions.

\begin{figure}[ht!]
  \centering
  \includegraphics[width=1.0\linewidth]{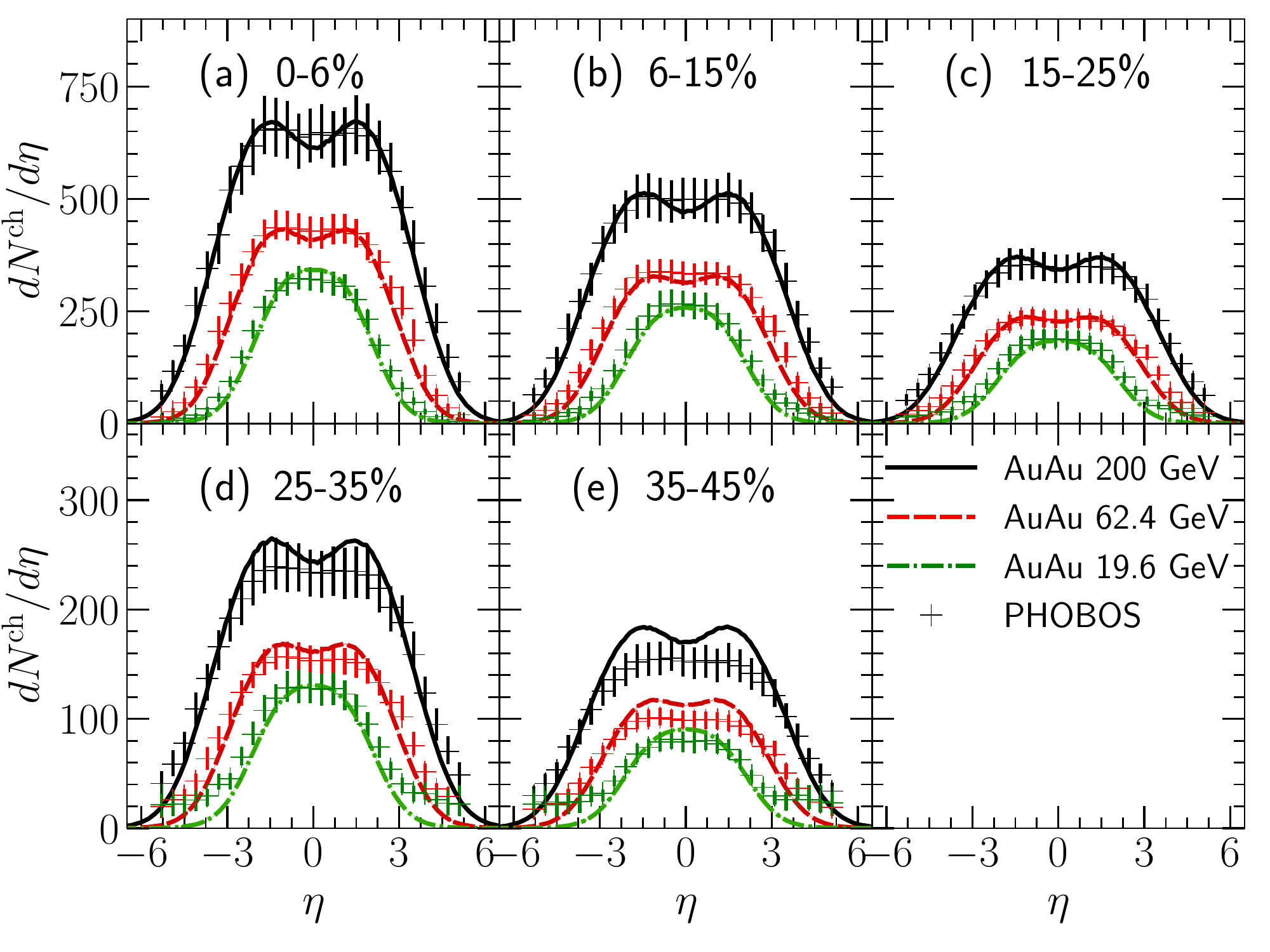}
  \caption{Pseudo-rapidity distributions of charged hadron production are compared with the PHOBOS measurements in Au+Au collisions at 19.6, 62.4, and 200 GeV \cite{Back:2005hs}.}
  \label{fig:dNchdeta1}
\end{figure}
%

Figure~\ref{fig:dNchdeta1} shows a comparison of charged hadron pseudo-rapidity distributions between our model and the PHOBOS measurements in Au+Au collisions at 19.6, 62.4, and 200 GeV \cite{Back:2005hs}. We only use the most central 0-6\% centrality data for model calibrations. Comparisons in the other four centrality bins test our model predictions. The dependence of particle productions on centrality is well reproduced with our 3D initial conditions.
We find that the rapidity boost to energy density profiles from the longitudinal momentum constraint in Eq.~(\ref{eq:ycm}) plays a crucial role in reproducing the measured centrality dependence of particle yields. The asymmetry between the two nuclear thickness functions at a given transverse position is larger in peripheral centrality bins compared to those in central collisions. The larger asymmetry at the edge of the transverse overlapping area boosts more energy density to forward and backward rapidity regions. This effect leads to a wider plateau of in the longitudinal profile for energy densities (illustrated in Fig.~\ref{fig:eprof}a). Without this local rapidity shift, the parameters in Table~\ref{table1} will overestimate mid-rapidity charged particle yields in peripheral Au+Au collisions.

\begin{figure}[ht!]
  \centering
  \includegraphics[width=1.0\linewidth]{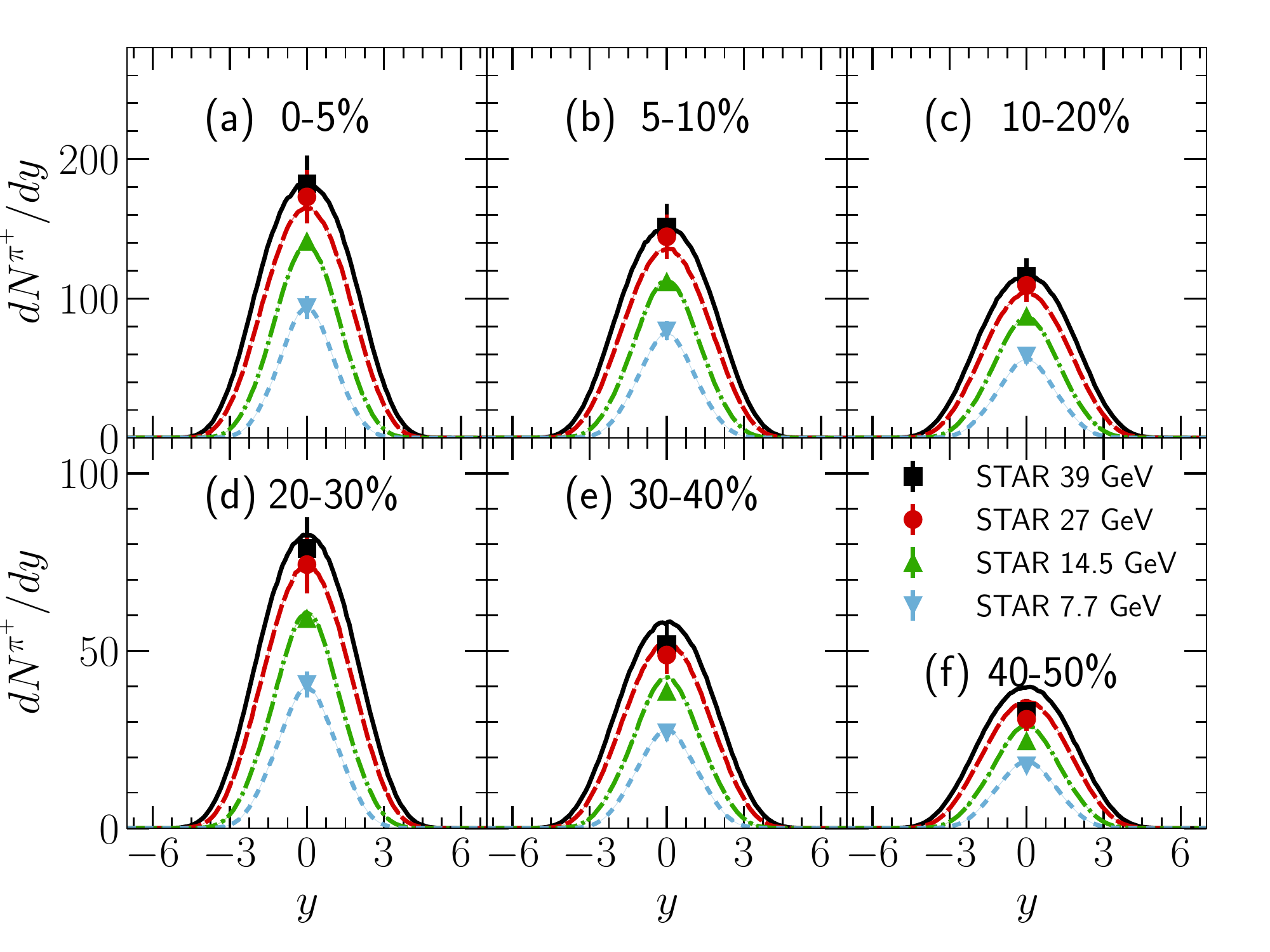}
  \caption{Rapidity distributions of $\pi^+$ are compared with the STAR measurements in Au+Au collisions at 7.7, 14.5, 27, and 39 GeV \cite{Adamczyk:2017iwn}.}
  \label{fig:dNchdeta2}
\end{figure}
%

At four additional collision energies in the RHIC BES program, measurements of particle rapidity distributions are still very limited. We present our model predictions for the rapidity distributions of final positively charged pions in Fig.~\ref{fig:dNchdeta2}. Because we utilize the total collision energy as a constraint in our model, we only need to adjust the plateau widths of the initial energy density profiles to match the measured $\pi^+$ yields in $\vert \Delta y \vert < 0.1$ \cite{Adamczyk:2017iwn}. Again, we only use the most central 0-5\% data for calibrations. Figure~\ref{fig:dNchdeta2} shows that the measured centrality dependence of particle production is well predicted by our model. Our predictions for particle rapidity distributions can be compared with future measurements from the RHIC BES phase II program.

\begin{figure}[ht!]
  \centering
  \includegraphics[width=1.0\linewidth]{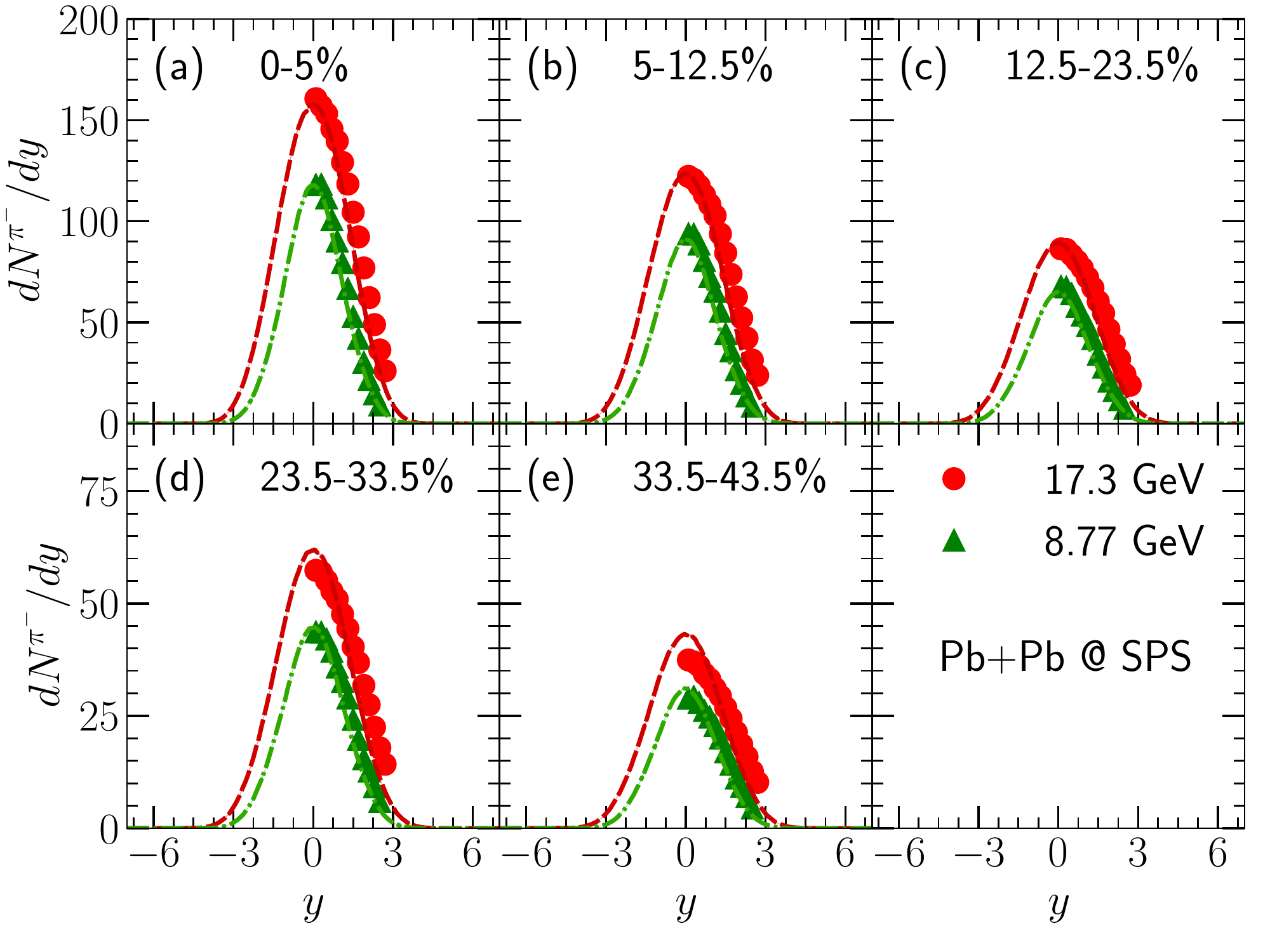}
  \caption{The rapidity distributions of $\pi^-$ production compare with the NA49 measurements in Pb+Pb collisions at $\sqrt{s_\mathrm{NN}} = 17.3$ and $8.77$ GeV \cite{Anticic:2012ay}.}
  \label{fig:dNchdeta3}
\end{figure}
%
We apply the same framework to simulate Pb+Pb collisions at two SPS collision energies. Rapidity distributions of $\pi^-$ in five centrality bins are compared with the NA49 measurements \cite{Anticic:2012ay} in Fig.~\ref{fig:dNchdeta3}. Once calibrated using measurements at the most central 0-5\% centrality, our model provides a reasonable description of the $\pi^-$ rapidity distributions in the rest four semi-peripheral centrality bins. The widths in $\pi^-$'s rapidity distributions at $\sqrt{s_\mathrm{NN}} = 17.3$\,GeV increase slightly faster in the experimental measurements than those in our model calculations.

\begin{figure}[ht!]
  \centering
  \includegraphics[width=0.9\linewidth]{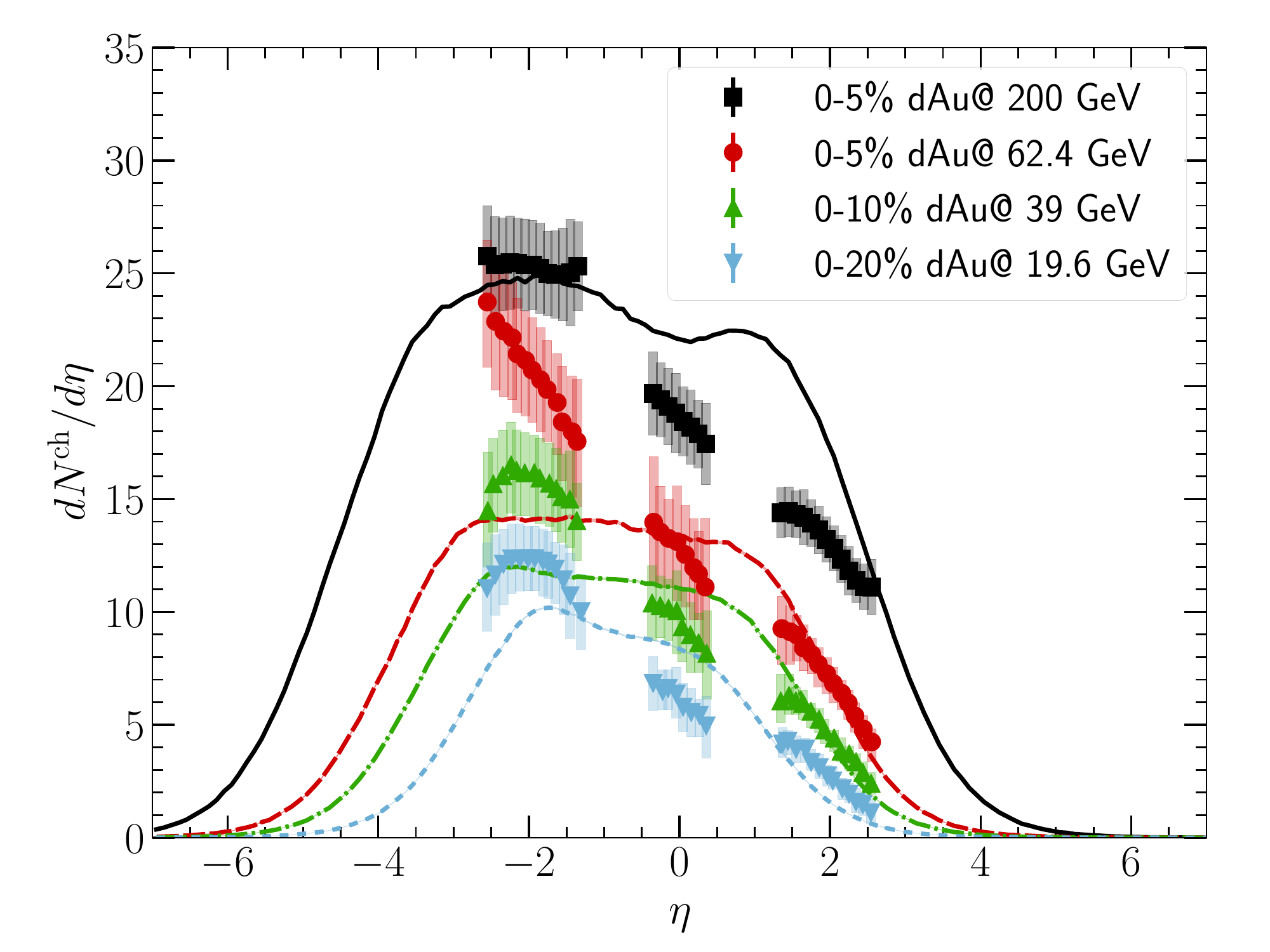}
  \caption{Model prediction of the charged hadron pseudo-rapidity distribution in central d+Au collisions. Comparisons are made with the PHENIX measurements for $\sqrt{s} = $19.6, 39, 62.4, and 200 GeV \cite{Adare:2018toe, Aidala:2017pup}.}
  \label{fig:dNchdeta_dAu}
\end{figure}
%
Last but not least, we test our model's predictive power by simulating highly asymmetric d+Au collisions at four collision energies. We use the same set of model parameters calibrated by central Au+Au collisions at the same collision energy. Figure~\ref{fig:dNchdeta_dAu} shows a comparison with the recent PHENIX measurements \cite{Adare:2018toe, Aidala:2017pup}. Our model qualitatively reproduces the collision energy dependence of charged hadron productions at mid-rapidity in d+Au collisions. However, the pseudo-rapidity distributions are somewhat too flat compared to the PHENIX measurements.  
These differences could come from the event-by-event fluctuations, which our current model has ignored.

\subsection{Baryon dynamics and net proton rapidity distributions}

Understanding the dynamics of baryon stopping is one of the main topics in the RHIC BES program. In this paper, we parameterize initial baryon distributions with Eqs.~(\ref{eq:nBprofr}) and (\ref{eq:nBprofl}). We do not intend to model the detailed baryon stopping mechanism, but rather to provide a parametric fit to all the existing heavy-ion measurements. This calibration makes sure all the collision systems are probing proper regions of the QCD phase diagram. To reduce the number of model parameters, we assume there is no baryon diffusion process during the hydrodynamic phase. The space-time rapidity profile for the initial baryon density is parametrized as the sum of two asymmetric Gaussian profiles for baryon charges carried by the projectile and target nucleus. We calibrated the parameters in Eqs.~(\ref{eq:nBprofr}) and (\ref{eq:nBprofl}) with the available net proton measurements in heavy-ion collisions.

\begin{figure}[ht!]
  \centering
  \includegraphics[width=1.0\linewidth]{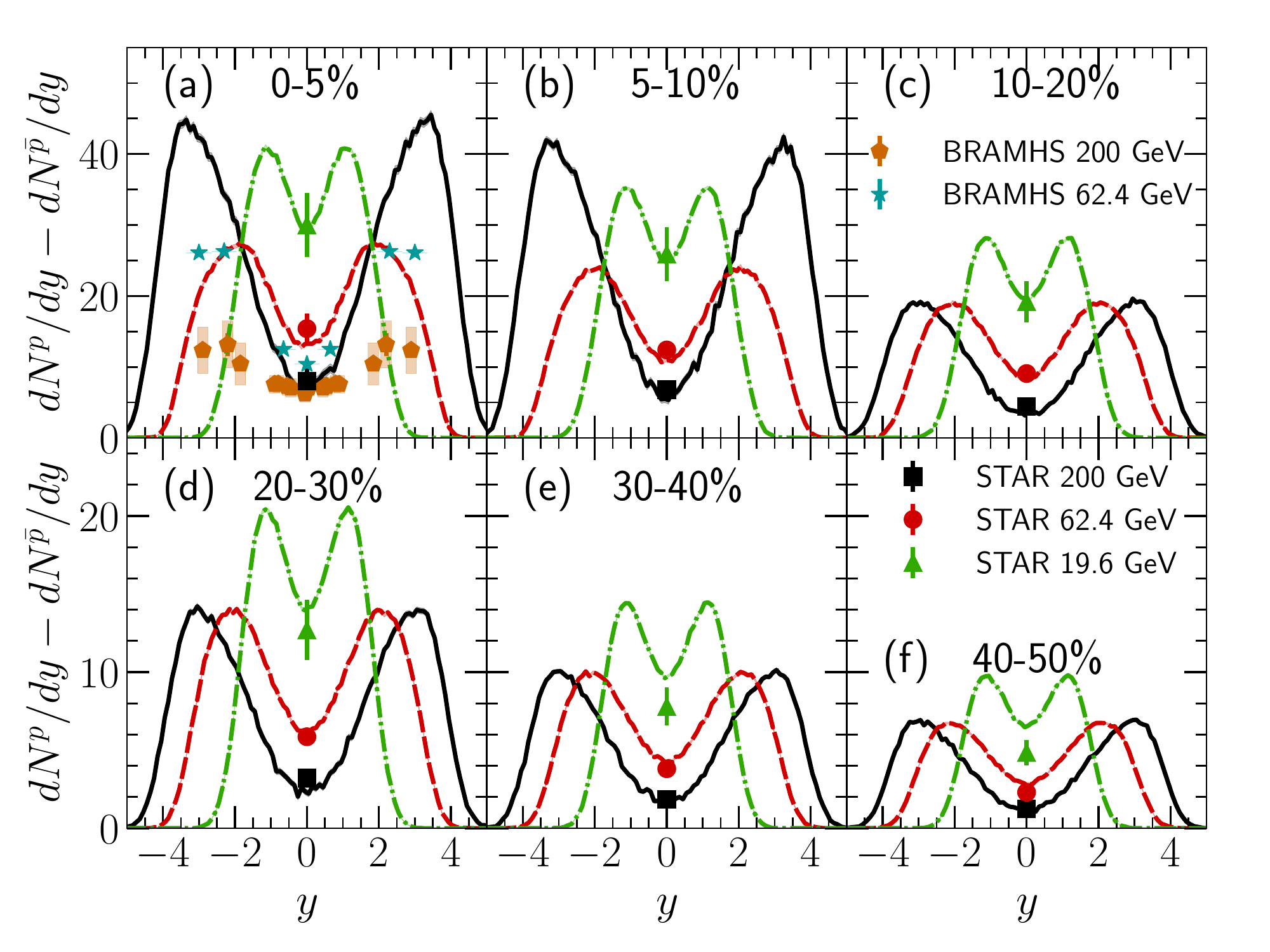}
  \caption{The net proton rapidity distributions in Au+Au collisions at 19.6, 62.4, and 200 GeV, compared with measurements from the BRAMHS and STAR collaborations \cite{Bearden:2003hx, Adamczyk:2017iwn, Abelev:2008ab}. Contributions from weak decays are included in the theoretical calculations.}
  \label{fig:netproton1}
\end{figure}

Figure~\ref{fig:netproton1} show net proton rapidity distributions in Au+Au collisions from 0-5\% to 40-50\% centrality bin at three collision energies. To compare with the RHIC measurements, we include feed-down contributions from weak decays from heavy resonance states to protons and anti-protons. Net proton distributions were measured by the BRAMHS Collaboration in the most central 0-5\% centrality in Au+Au collisions at 200 and 62.4 GeV \cite{Bearden:2003hx}. The mid-rapidity measurements by the STAR Collaboration \cite{Abelev:2008ab} is consistent with the BRAMHS results. By adjusting the model parameters in Table~\ref{table1}, we can reasonably match the BRAMHS measurements near mid-rapidity. For 19.6 GeV, our model parameters are only guided by the STAR data at $y=0$. Nevertheless, our model predictions in Figs.~\ref{fig:netproton1}b-f give good descriptions of the STAR measurements in semi-peripheral Au+Au collisions.

\begin{figure}[ht!]
  \centering
  \includegraphics[width=1.0\linewidth]{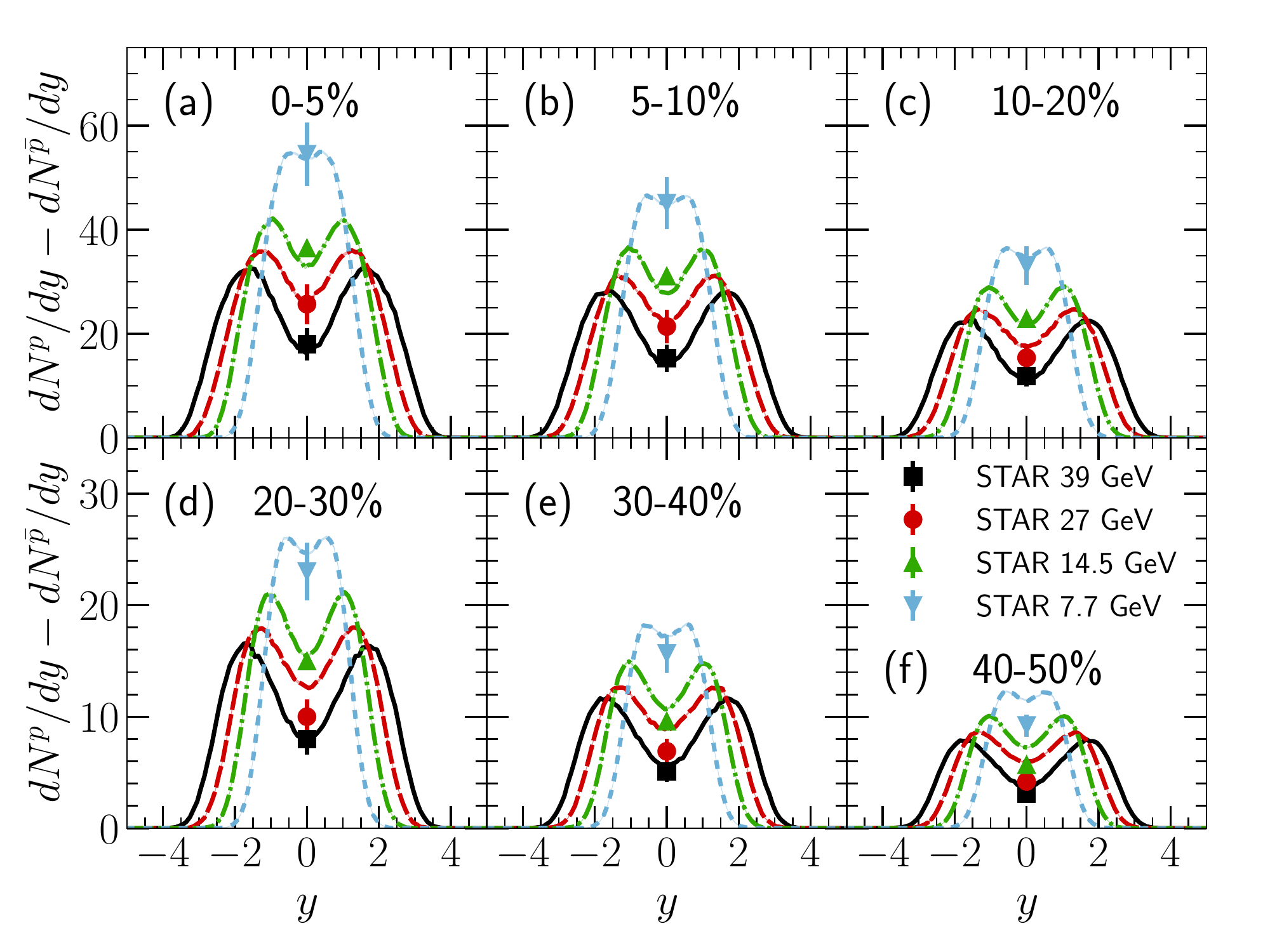}
  \caption{The net proton rapidity distributions in Au+Au collisions at 7.7, 14.5, 27, and 39 GeV, compared with measurements from the STAR collaborations at mid-rapidity \cite{Adamczyk:2017iwn}. Contributions from weak decays are included in the theoretical calculations.}
  \label{fig:netproton2}
\end{figure}
%
At additional four collision energies in the RHIC BES program, rapidity distributions of net-proton distributions have not been measured yet. We adjust our model parameters to match the measured net proton yields at mid-rapidity in the central 0-5\% Au+Au collisions shown in Fig.~\ref{fig:netproton2}. Our results in the rest five centrality bins can be considered as model predictions. We slightly overestimated the mid-rapidity net proton yield in the 40-50\% centrality bin, which hints for a weaker baryon stopping in peripheral collisions than central ones.

\begin{figure}[ht!]
  \centering
  \includegraphics[width=1.0\linewidth]{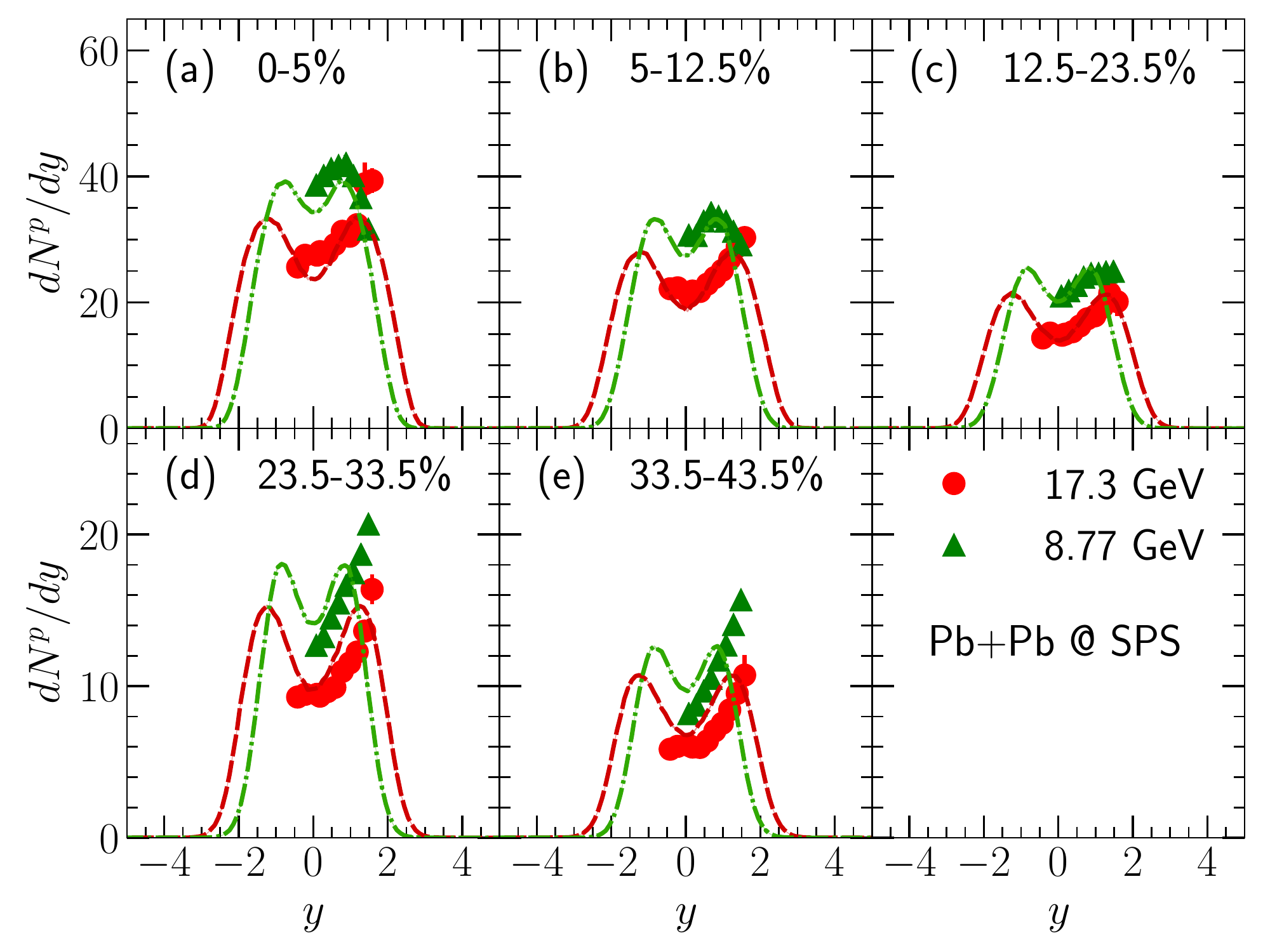}
  \caption{The proton rapidity distributions in Pb+Pb collisions at $\sqrt{s_\mathrm{NN}} = 17.3$ and $8.77$ GeV, compared with measurements from the NA49 collaborations \cite{Anticic:2010mp}.}
  \label{fig:netproton3}
\end{figure}
%
Finally, we study proton rapidity distributions in Pb+Pb collisions at CERN SPS collision energies in Fig.~\ref{fig:netproton3}. The NA49 Collaboration measured proton rapidity distributions at center-of-mass energies $\sqrt{s_\mathrm{NN}} = 17.3$ and $8.77$ GeV for five centrality bins \cite{Anticic:2010mp}. Our calculations give an overall good description to the NA49 measurements. 
Figures ~\ref{fig:netproton3}d-e show that the proton yields keep increasing in forward rapidity regions in semi-peripheral collisions, which is not seen in our calculations. This qualitative difference may suggest that the experimental data has a contamination of protons from the spectators at forward rapidity regions. Our model parameters used to calibrate Pb+Pb collisions at SPS collision energies are consistent with those used in Au+Au collisions at similar RHIC BES energies.

\subsection{Rapidity dependent anisotropic flow}

The rapidity distributions of anisotropic flow observables can elucidate the coupling between transverse and longitudinal dynamics in heavy-ion collisions. In this section, we benchmark our results with event-averaged initial conditions, which contain high degrees of symmetry. These flow results can be viewed as a baseline for future comparisons with those from more realistic event-by-event simulations.

\begin{figure}[ht!]
    \centering
    \includegraphics[width=1.0\linewidth]{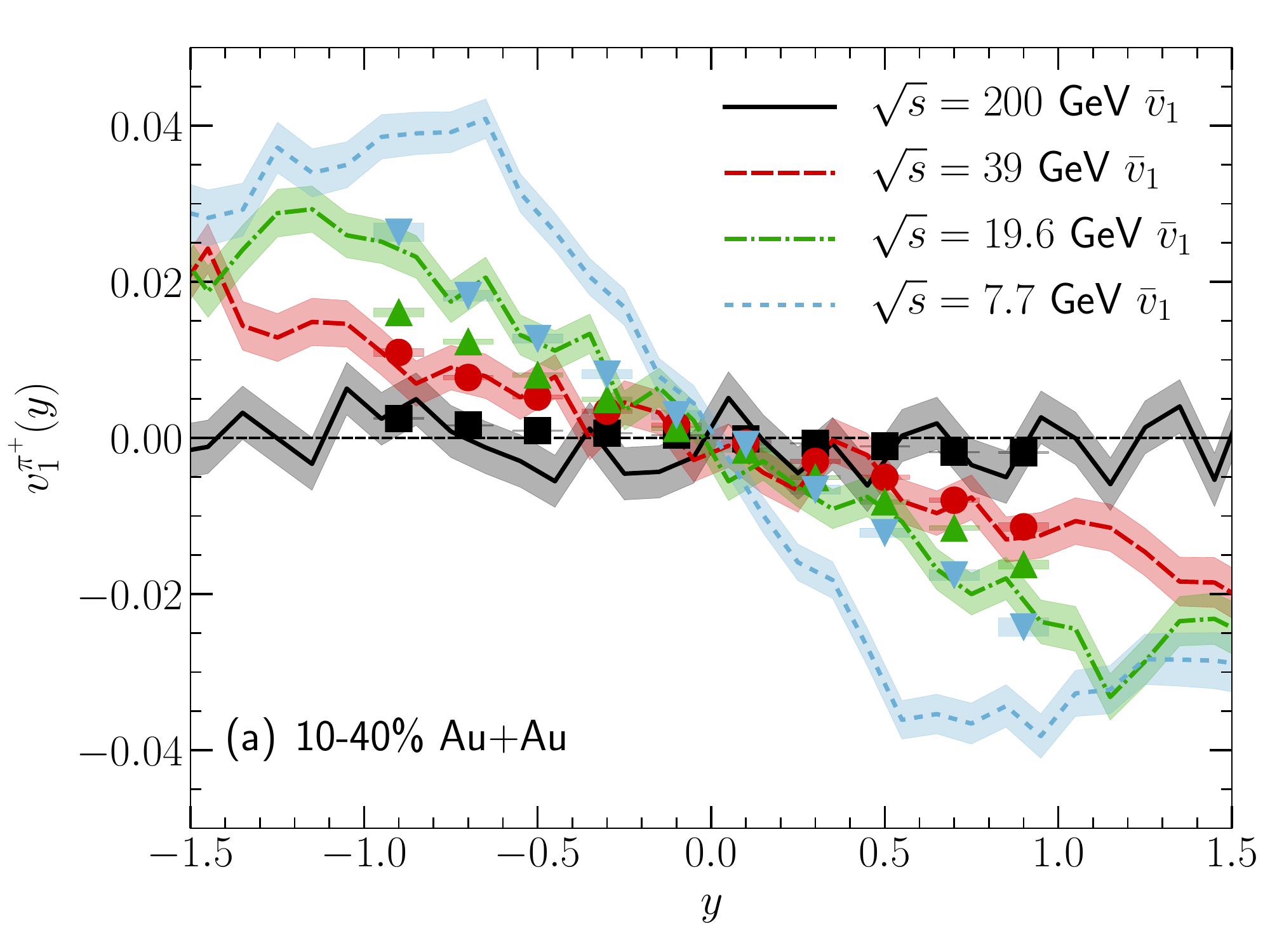}
    \includegraphics[width=1.0\linewidth]{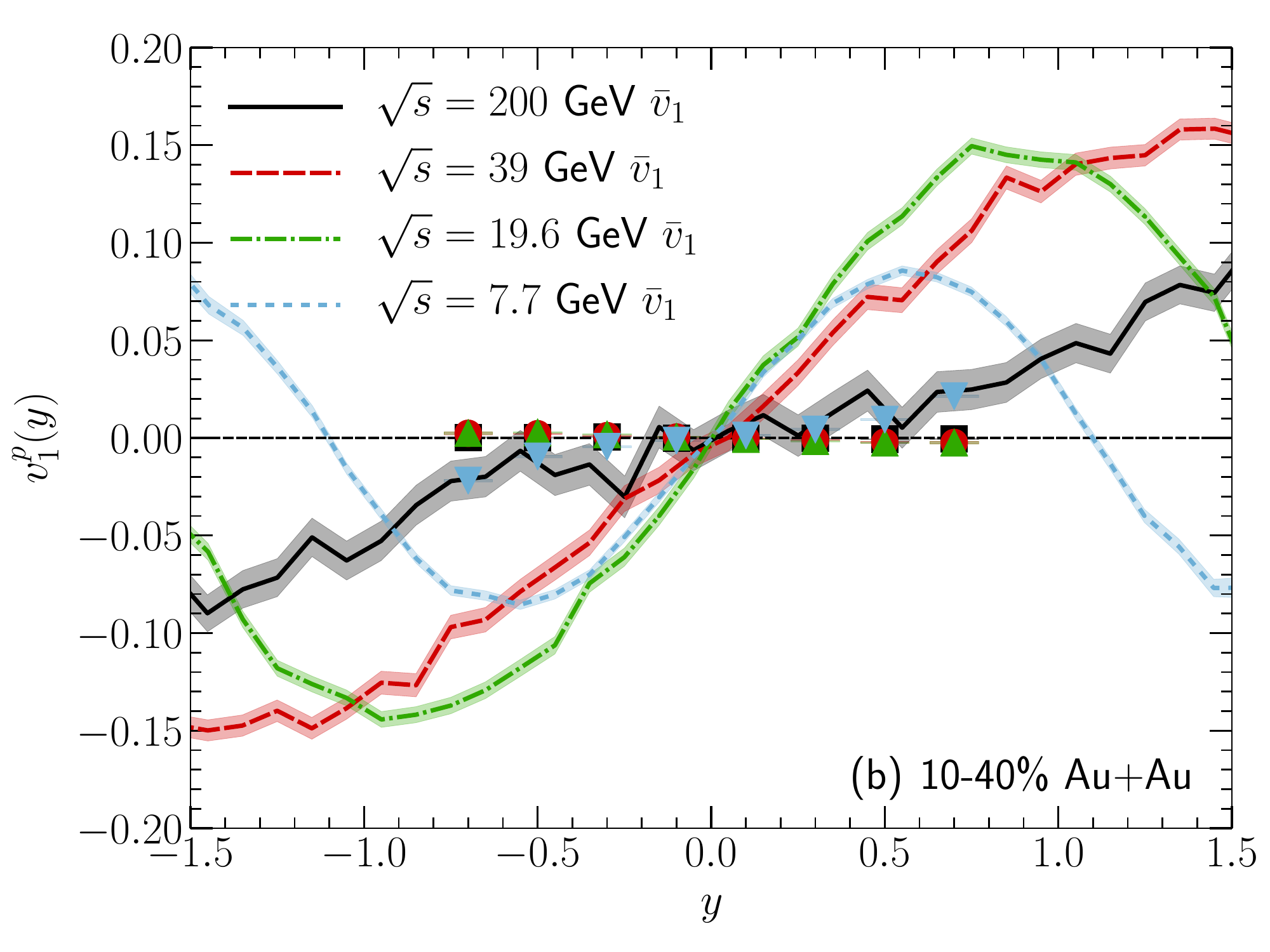}
    \caption{Identified particle directed flow coefficients for $\pi^+$ (a) and protons (b) as functions of rapidity in 10-40\% Au+Au collisions at $\sqrt{s_\mathrm{NN}} = $200, 39, 19.6, and 7.7 GeV. Comparisons are made with the STAR measurements \cite{Adamczyk:2014ipa}. }
    \label{fig:v1eta}
\end{figure}
Figure~\ref{fig:v1eta} shows the rapidity dependent identified particle directed flow coefficients at four collision energies. The slope of $\pi^+$'s directed flow $dv_1^{\pi^+}/dy\vert_{y=0}$ is negative at mid-rapidity, and its magnitude is larger at lower collision energy. Our simulation shows that the $v_1(y)$ of $\pi^+$ follows the dipole deformation of the initial energy density profile in Fig.~\ref{fig:eccn_rap}a. This correlation reflects that the directed flow of $\pi^+$ is mainly driven by local pressure gradients, similar to the dynamics of elliptic flow. In the meantime, the proton $v_1(y)$ has an opposite sign compared to that of $\pi^+$ at all collision energies. This is because proton $v_1(y)$ receives additional contributions from initial decelerated baryons other than thermal emissions from fluid cells. These decelerate baryons carry a directed flow pointing to the spectators' direction in the transverse plane. 

Our $\pi^+$'s directed flow agree reasonably with the STAR measurements for $\sqrt{s} \ge 39$\,GeV \cite{Adamczyk:2014ipa}. They start to overestimate the experimental data at lower collision energies.
Our model overestimated proton's directed flow by more than an order of magnitude. The slopes of proton $v_1$ are all positive for all collision energies 
This failure suggests that our ansatz for initial baryon distributions in Eqs.~(\ref{eq:nBprofr}) and (\ref{eq:nBprofl}) is too simple. More realistic modeling of the initial baryon stopping mechanism is needed to understand the RHIC measurements.

\begin{figure}[h!]
    \centering
    \includegraphics[width=1.0\linewidth]{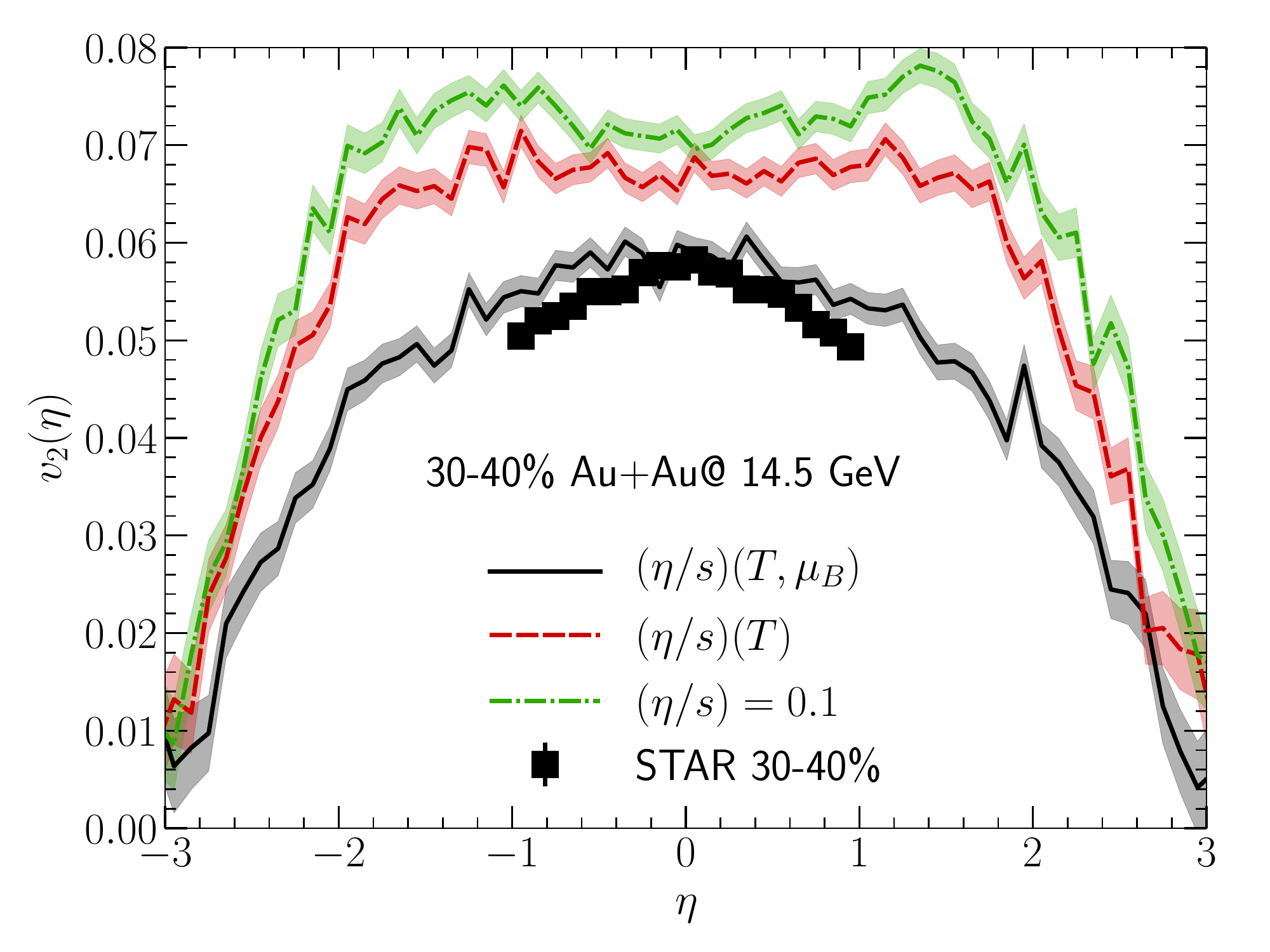}
    \caption{Pseudo-rapidity distributions of charged hadron elliptic flow coefficients in 30-40\% Au+Au collisions at $\sqrt{s_\mathrm{NN}} = $14.5 GeV with different $(\eta/s)(T, \mu_B)$. The charged hadron $v_2(\eta)$ coefficients are integrated from $p_T = 0.2$ to 3 GeV and compared with the STAR measurements \cite{Adam:2019dkq}. }
    \label{fig:v2eta}
\end{figure}

Figure~\ref{fig:v2eta} further shows pseudo-rapidity distributions of charged hadron elliptic flow at 14.5 GeV. Our full hybrid simulations with event-averaged initial conditions (the solid black line) give a flatter distribution of $v_2(\eta)$ near mid-rapidity compared to the STAR measurements \cite{Adam:2019dkq}. Our $v_2(\eta)$ distribution is a result of the cancellation between large effective shear viscosity in large $\mu_B$ regions and large initial eccentricity $\varepsilon_2$ in the forward and backward rapidity regions (see Fig.~\ref{fig:eccn_rap}b). Simulations with constant or only temperature-dependent specific shear viscosity give larger charged hadron $v_2$ at $\eta \sim 2$ compared to its value at mid-rapidity in Fig.~\ref{fig:v2eta}. This discrepancy between our full results and the STAR measurements suggests that flow longitudinal decorrelations rooted from event-by-event fluctuations are essential to understand this observable \cite{Behera:2020mol}. We devote the extension to event-by-event simulations and studying longitudinal flow fluctuations to a future work.

\subsection{Study QGP transport properties with the $\sqrt{s}$-dependent transverse dynamics} \label{sec:results:transflow}

Hydrodynamic flow boosts the thermally emitted hadrons, which results in increasing their mean transverse momenta. In this work, we neglect bulk viscous effects in the QGP evolution and adjust the starting time of hydrodynamics at every collision energy so that identified particles' $\langle p_T \rangle$ match to the STAR measurements in 0-5\% Au+Au collisions \cite{Abelev:2008ab, Adamczyk:2017hdl, Adam:2019dkq}.

\begin{figure}[ht!]
  \centering
       \includegraphics[width=0.95\linewidth]{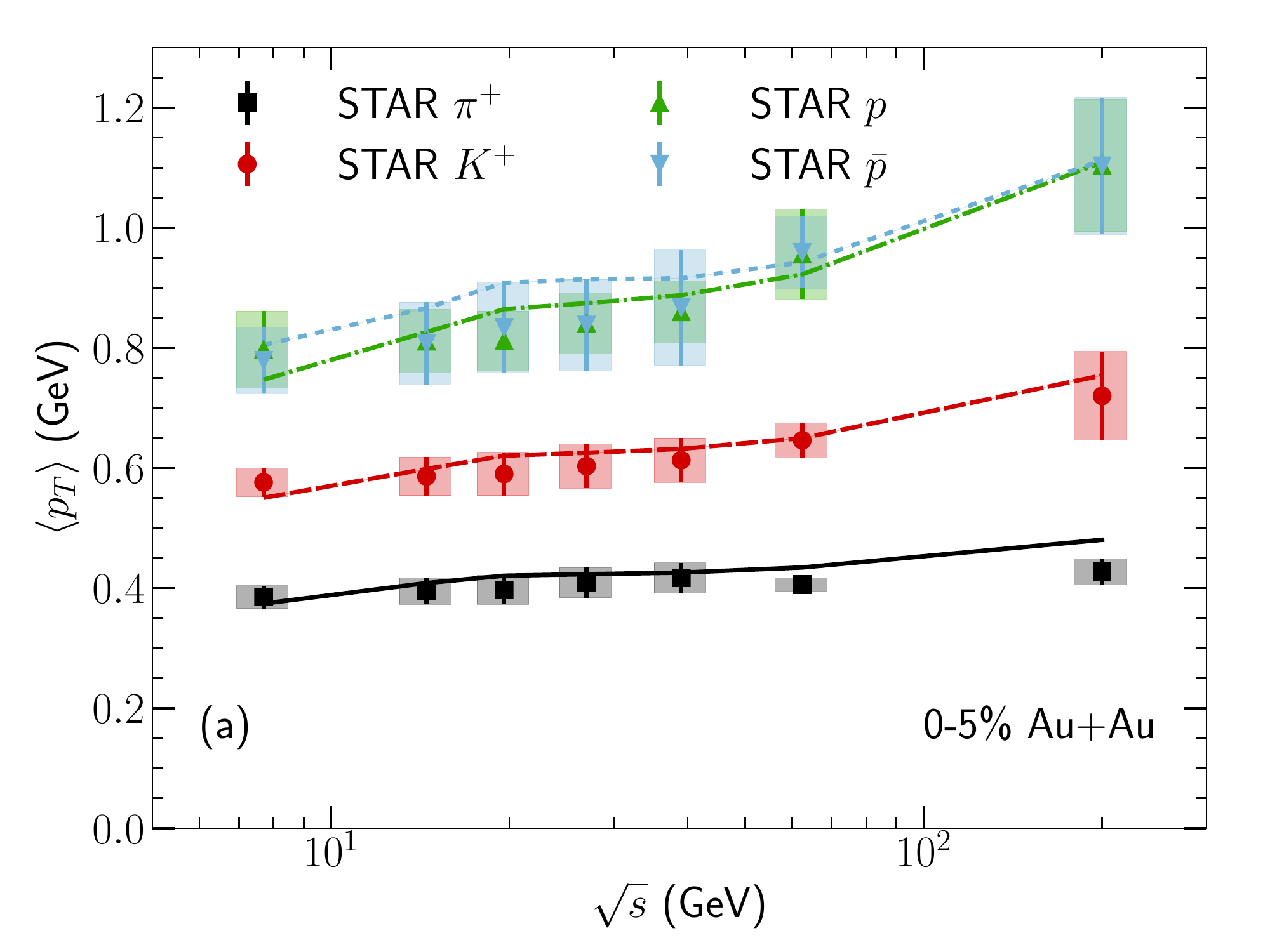}
       \includegraphics[width=0.95\linewidth]{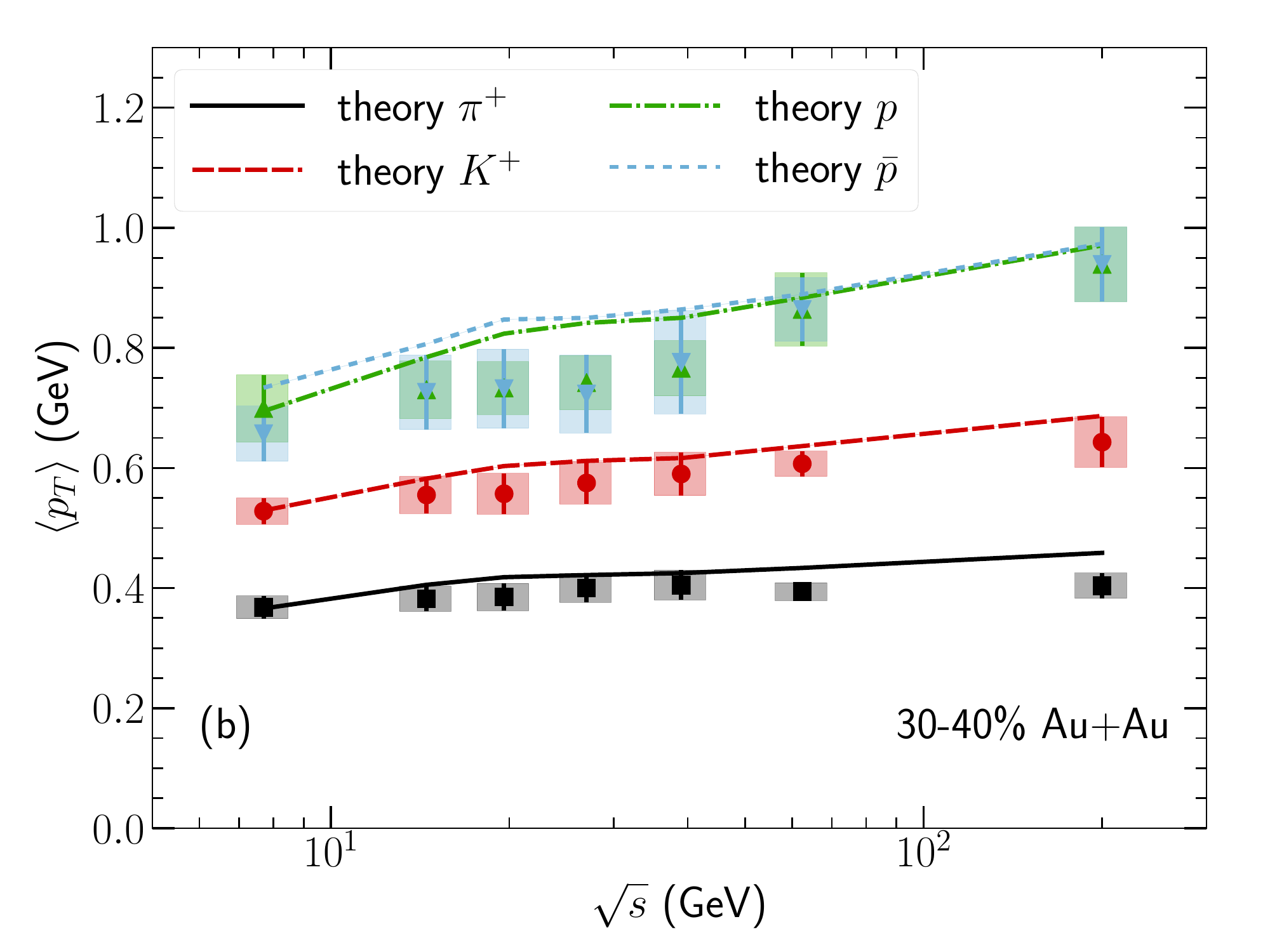}
  \caption{Collision energy dependence of the identified particle averaged transverse momentum $\langle p_T \rangle$ in Au+Au collisions at 0-5\% (a) and 30-40\% (b) centrality bins. Theoretical results are compared with the STAR measurements in Refs.~\cite{Abelev:2008ab, Adamczyk:2017hdl, Adam:2019dkq}. }
  \label{fig:meanpT}
\end{figure}

Figure~\ref{fig:meanpT} shows that the identified particle mean $p_T$ as a function of collision energy for 0-5\% and 30-40\% centrality bins. Our theoretical calculations can quantitatively reproduce the STAR measurements and capture the collision energy dependence of $\langle p_T \rangle$. It is clear from the proton's mean $p_T$ that systems at higher collision energy develop stronger radial flow because their hydrodynamic phases are longer. Our calculation suggests that the mean $p_T$ of anti-protons is larger than that of protons in all collision energies. This difference increases with the net baryon chemical potential at low collision energies. The systematic uncertainties in the current proton and anti-proton measurements are still too large to distinguish them from each other.

Please note that our results ignore any pre-equilibrium dynamics before the hydrodynamic starting time $\tau_0$ listed in Table.~\ref{table1}. Any pre-hydrodynamic evolution \cite{Liu:2015nwa,Kurkela:2018wud,Kurkela:2018vqr,Gale:2020xlg} will generate transverse flow during $\tau = 0^+ - \tau_0$, which will result in stronger transverse flow at the freeze-out and larger particle mean $p_T$. This strong radial flow needs to be tamed by bulk viscous effects during hydrodynamic evolution \cite{Ryu:2015vwa, Ryu:2017qzn}. Therefore, the fact that our current results can reproduce the STAR mean $p_T$ measurements suggest a non-zero QGP bulk viscosity at finite densities. 

In high energy collisions, the mid-rapidity elliptic flow coefficient of charged hadrons is driven by the fireball's elliptic deformation $\varepsilon_2(\eta_s)$ \cite{Li:2019eni, Franco:2019ihq}.
Although the mid-rapidity $\varepsilon^\mathrm{3D}_2(\eta_s = 0)$ for a given centrality class of Au+Au collisions only varies by few percents from 200 to 7.7 GeV, the plateau of $\varepsilon_2(\eta_s)$ shrinks quickly in space-time rapidity as the collision energy goes down (see Fig.~\ref{fig:eccn_rap}b). The averaged shape of the fireball in $\vert \eta_s \vert < 1$ is more eccentric at a lower collision energy. This larger averaged initial eccentricity is an effect from the breaking of boost-invariance in full 3D simulations. The detailed 3D structure of $\varepsilon^\mathrm{3D}_2(\eta_s)$ is important for the elliptic flow development during hydrodynamic simulations.

\begin{figure}[ht!]
  \centering
       \includegraphics[width=0.95\linewidth]{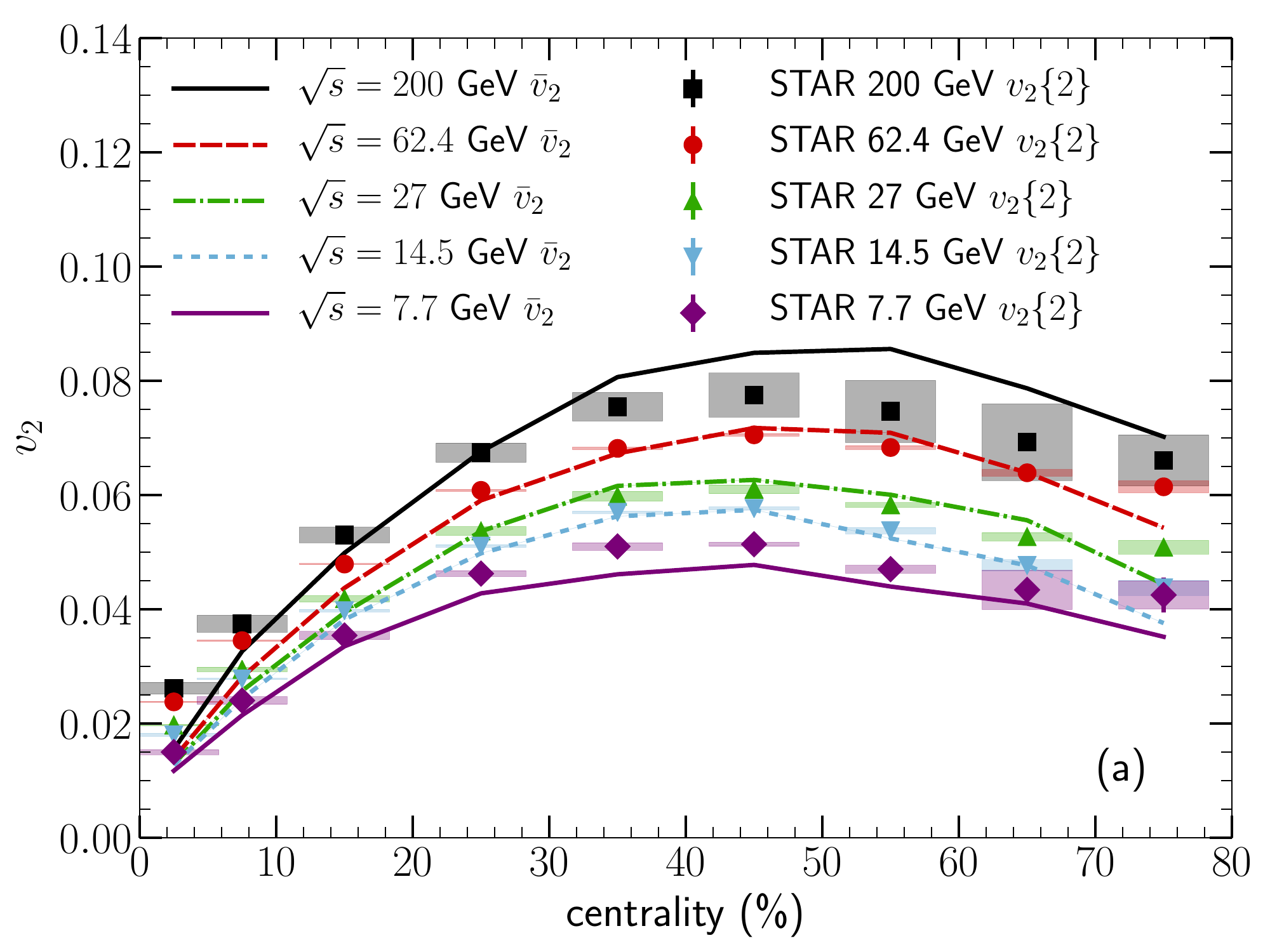} 
       \includegraphics[width=0.95\linewidth]{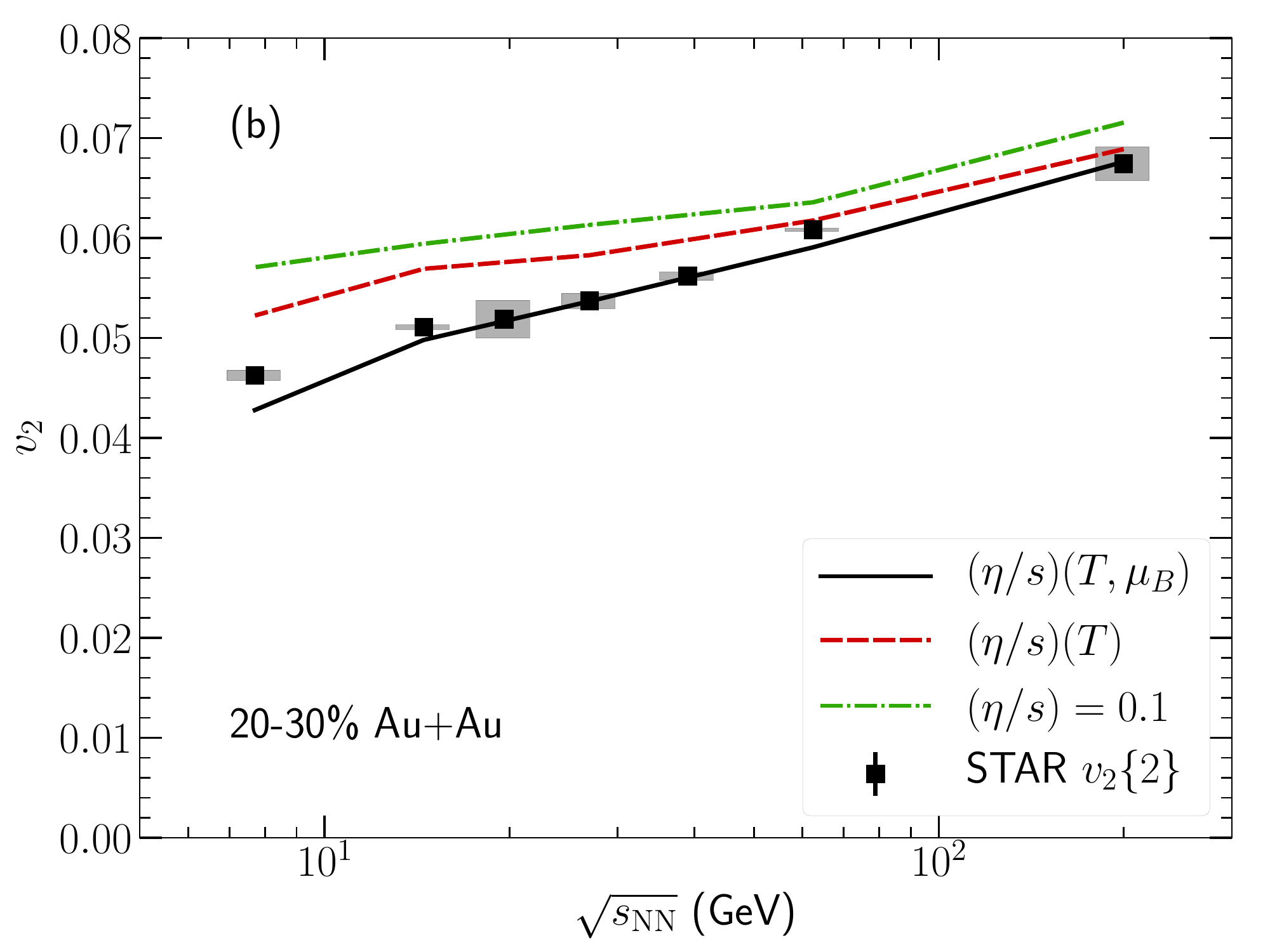}
  \caption{The dependence of elliptic flow coefficients on collision centrality (a) and collision energy (b) in Au+Au collisions from 7.7 to 200 GeV. Theoretical results are compared with STAR measurements using the 2-particle cumulant method \cite{Adamczyk:2017hdl}.}
  \label{fig:v2}
\end{figure}
%

Figure~\ref{fig:v2}a show our model calibration on the STAR elliptic flow measurements with the specific shear viscosity $(\eta/s)(T, \mu_B)$ shown in Fig.~\ref{fig:vis}.
Please note that our simulations neglect event-by-event fluctuations and bulk viscous effects. Although these two factors give opposite contribution to the elliptic flow coefficient, we expect $\mathcal{O}(20\%)$ theoretical uncertainty in our calculations. Therefore, our following discussion will focus on qualitative features in the model-to-data comparisons.
In order to reproduce the collision energy dependence of the STAR elliptic flow measurements, we require a larger the QGP specific shear viscosity in the low $T$ and large $\mu_B$ regions. Our conclusion is inline with the finding in Ref.~\cite{Karpenko:2015xea}, in which the effective $\eta/s(\sqrt{s})$ is larger at a lower collision energy.
We further study individual effects from $T$ and $\mu_B$ dependence in $\eta/s$ on the collision energy dependent elliptic flow coefficients in Fig.~\ref{fig:v2}b. The $\mu_B$ dependence in $\eta/s$ give a large contribution to the suppression of charged hadron $v_2(\sqrt{s})$ at low collision energies, at which more baryons are stopped at mid-rapidity.
Full 3D simulations are essential to extract the $T$ and $\mu_B$ dependent $\eta/s(T, \mu_B)$. If one assumes systems are boost-invariant at all these collision energies, one would need a roughly constant $\eta/s$ to reproduce the collision energy dependence of charged hadron $v_2$ \cite{Shen:2012vn}.

Our results in Fig.~\ref{fig:v2eta} and Fig.~\ref{fig:v2} have demonstrated that full 3D dynamics of heavy-ion collisions can set strong constraints on the temperature and chemical potential dependent QGP specific shear viscosity $(\eta/s)(T, \mu_B)$ by using the elliptic flow measurements as functions of particle rapidity and collision energy in the RHIC BES program.

\section{Conclusions}
\label{sec:conclusion}

In this work, we present a simple way to construct 3D initial conditions for relativistic heavy-ion collisions at any energy. The key ingredient is that our formulation ensures local energy and momentum conservation at all transverse position $(x, y)$ of the system. The conservation laws, together with our choice of the longitudinal energy density profile in Eq.~(\ref{eq:eprof}), result in a non-trivial $\sqrt{T_A T_B}$ scaling for the local energy density $e(x, y)$ at high energies. This scaling qualitatively agrees with the phenomenological constrained results from state-of-the-art Bayesian statistical analyses \cite{Bernhard:2016tnd, Bernhard:2019bmu}. Therefore, our work provides an alternative physical interpretation for this type of scaling, in addition to saturation physics \cite{Moreland:2014oya}. From our model's point of view, the fact that $e \propto \sqrt{T_A T_B}$ is a consequence of longitudinal momentum conservation and the assumption of a flux-tube like longitudinal profile for the local energy density.

We systematically study this type of 3D initial conditions using viscous hydrodynamics + hadronic transport hybrid framework. We calibrate this framework with charged hadron and proton rapidity distributions in most central heavy-ion collisions in the RHIC BES program and at CERN SPS energies. By fixing all the parameters, we test model predictions in semi-peripheral Au+Au and Pb+Pb collisions as well as in asymmetric d+Au collisions. The proposed 3D initial conditions can quantitatively reproduce particle rapidity distributions measured in different centrality bins and in d+Au collisions. These successful predictions support that imposing local energy-momentum conservation plays a critical role in understanding the evolution charged hadron pseudo-rapidity distributions as a function of collision geometry in heavy-ion collisions.
In the meantime, our model with event-averaged initial conditions failed to describe the rapidity dependent $v_1(y)$ of pions and protons. These discrepancies indicate that a realistic baryon stopping mechanism and event-by-event fluctuations play critical roles in understanding the directed flow observables.

We further study the collision energy dependence of flow observables at mid-rapidity. Our model without any pre-equilibrium dynamics before $\tau = \tau_0$ can reproduce the measured mean $p_T$ of identified particles. This comparison hints that a non-zero QGP bulk viscosity is essential in the more realistic heavy-ion simulations at finite densities.
We demonstrate that the charged hadrons $v_2(\sqrt{s})$ in the RHIC BES program has strong constraining power on the temperature and $\mu_B$ dependence of the QGP specific shear viscosity.

The calibrated simulations presented here serve as a baseline result for the bulk dynamics in the RHIC BES program. Future comparisons with more realistic simulations will quantitatively address the importance of event-by-event fluctuations and pre-equilibrium evolution on flow observables at different collision energies. Last but not least, our calibrated medium evolution in this work can be directly used to study the evolution of critical fluctuations \cite{Stephanov:2017ghc, Rajagopal:2019xwg, An:2019csj}, the effect of strong electric magnetic fields \cite{Gursoy:2018yai, Gursoy:2020jso}, and the QGP electromagnetic radiation in a baryon rich environment \cite{Gale:2018vuh, Vujanovic:2019yih}.

All software used in this work are open source:
\begin{itemize}
    \item \texttt{iEBE-MUSIC}  \cite{iEBEMUSIC}: A fully integrated numerical framework to automate hybrid simulations for relativistic heavy-ion collisions.
    \item \texttt{superMC} \cite{superMC, Shen:2014vra}: An initial condition generator based on Monte-Carlo Glauber model.
    \item \texttt{MUSIC} \cite{MUSIC, Schenke:2010nt, Schenke:2011bn, Paquet:2015lta, Denicol:2018wdp}: A (3+1)D relativistic viscous hydrodynamics
    \item \texttt{iSS} \cite{iSS, Shen:2014vra}: A Monte-Carlo particlization module based on the Cooper-Frye freeze-out procedure
    \item \texttt{UrQMD} \cite{UrQMD, Bass:1998ca, Bleicher:1999xi}: A standard hadronic transport model 
    \item \texttt{Hadronic\_afterburner\_toolkit} \cite{Toolkit}: A particle spectra and flow analysis code package
\end{itemize}
The event averaged initial conditions, hydrodynamic hyper-surfaces, and final experimental observables can be downloaded from the following link \cite{Publicresults}.

\acknowledgments

The authors thank A. Monnai for giving supports on equation of state and D. Oliinychenko for fruitful discussions.
This work is supported under DOE Contract No. DE-SC0013460 and in part by the U.S. Department of Energy, Office of Science, Office of Nuclear Physics, within the framework of the Beam Energy Scan Theory (BEST) Topical Collaboration.
S. A. acknowledges scholarship supports from the Department of Physics, Jazan Univeristy, Jazan, Kingdom of Saudi Arabia. 
This research used resources of the National Energy Research Scientific Computing Center, which is supported by the Office of Science of the U.S. Department of Energy under Contract No. DE-AC02-05CH11231 and resources of the high performance computing services at Wayne State University.

\bibliography{ref}

\end{document}